\documentclass[twocolumn,amsmath,amssymb,prd]{revtex4-1}

\usepackage{graphicx}

\def\al{\alpha}
\def\be{\beta}
\def\ga{\gamma}
\def\de{\delta}
\def\ep{\epsilon}

\def\et{\eta}

\def\ka{\kappa}
\def\la{\lambda}

\def\rh{\rho}

\def\si{\sigma}

\def\ps{\psi}

\def\La{\Lambda}

\def\half{{\textstyle{\frac 1 2}}}
\def\prt{\partial}

\newcommand{\beq}{\begin{equation}}
\newcommand{\eeq}{\end{equation}}
\newcommand{\bea}{\begin{eqnarray}}
\newcommand{\eea}{\end{eqnarray}}
\newcommand{\rf}[1]{(\ref{#1})}
\newcommand{\nn}{\nonumber}

\def\etal{{\it et al.}}

\def\pt#1{\phantom{#1}}
\def\ol#1{\overline{#1}{}}

\def\nnn{N^\prime}
\def\wth{{\widetilde h}}
\def\wtn{{\widetilde n}}
\def\wto{{\widetilde O}}
\def\wtnn{{\widetilde N}}
\def\wtt{{\widetilde T}}

\def\vb#1#2{e_{#1}^{{\pt{#1}}#2}}

\def\aon{\rh_1}
\def\ato{\rh_2}
\def\att{\rh_3}
\def\atho{\rh_4}
\def\atht{\rh_5}
\def\afo{\rh_6}
\def\afio{\rh_7}
\def\afit{\rh_8}
\def\asio{\rh_9}
\def\asit{\rh_{10}}
\def\aseo{\rh_{11}}
\def\don{\rh_{12}}
\def\dtw{\rh_{13}}
\def\dth{\rh_{14}}
\def\dfo{\rh_{15}}
\def\dfi{\rh_{16}}
\def\bon{\si_1}
\def\btw{\si_2}
\def\bth{\si_3}
\def\bfo{\si_4}
\def\bft{\si_5}
\def\bfi{\si_6}

\begin{document}

\title{
Lorentz symmetry in ghost-free massive gravity}

\author{
V.\ Alan Kosteleck\'y$^1$ and Robertus Potting$^{2,3}$}

\affiliation{
$^1$Physics Department, Indiana University, 
Bloomington, Indiana 47405, USA}

\affiliation{
$^2$Universidade do Algarve, Faculdade de Ci\^encias e Tecnologia, 
8005-139 Faro, Portugal}

\affiliation{
$^3$CENTRA, IST, Universidade de Lisboa, 
Avenida Rovisco Pais, 1049-001 Lisboa, Portugal}

\date{August 2021} 

\begin{abstract}

The role of Lorentz symmetry in ghost-free massive gravity is studied,
emphasizing features emerging in approximately Minkowski spacetime.
The static extrema and saddle points of the potential are determined
and their Lorentz properties identified.
Solutions preserving Lorentz invariance and ones breaking
four of the six Lorentz generators are constructed.
Locally, globally, and absolutely stable Lorentz-invariant extrema
are found to exist for certain parameter ranges of the potential.
Gravitational waves in the linearized 
theory
are investigated.
Deviations of the fiducial metric from the Minkowski metric
are shown to lead to pentarefringence of the five wave polarizations,
which can include superluminal modes and subluminal modes
with negative energies in certain observer frames.
The Newton limit of ghost-free massive gravity is explored.
The propagator is constructed and used to obtain 
the gravitational potential energy between two point masses.
The result extends the Fierz-Pauli limit to include corrections 
generically breaking both rotation and boost invariance.

\end{abstract}

\maketitle

\section{Introduction}
\label{Introduction}

The foundational symmetries of General Relativity (GR) 
include local Lorentz invariance and diffeomorphism invariance,
which exclude the graviton from acquiring a mass.
In linearized gravity,
a massive graviton propagating on Minkowski spacetime
can be described by the Fierz-Pauli term 
\cite{fp39}.
However,
in the limit of vanishing mass,
the Fierz-Pauli term is discontinous to linearized GR
\cite{vvd70,vz70},
as in addition to the usual graviton modes
it yields a massless scalar mode,
the Boulware-Deser ghost
\cite{bd72}.
A theory of massive gravity reducing to GR in the massless limit
requires instead a nonlinear completion
\cite{av72}.
This can be constructed using a special nonlinear potential
that eliminates the ghost mode 
due to the appearance of an extra second-class constraint
\cite{drgt11}.
The action for massive gravity
can be formulated using two metrics,
a dynamical metric $g_{\mu\nu}$
and a nondynamical fiducial metric $f_{\mu\nu}$
\cite{hr11},
combined in a five-parameter quartic potential
that removes the two ghost degrees of freedom
\cite{hr1112}.
The theory of massive gravity and its developments,
including bimetric and multimetric versions,
are reviewed in Refs.\ \cite{kh11,dr14,ss16}.

The nondynamical fiducial metric $f_{\mu\nu}$
in ghost-free massive gravity 
can be viewed as a prescribed background field
that explicitly breaks the spacetime symmetries of GR in a special way,
thereby permitting a nonzero graviton mass with no ghost mode.
Astrophysical constraints require the graviton mass 
to be less than $10^{-38}$ GeV
\cite{gn10},
so a nonzero value would represent 
a phenomenologically tiny deviation from GR.
In a broad context,
small deviations from a specified theory can be described
in a model-independent way using effective field theory
\cite{sw09}.
The general effective field theory
containing all terms breaking the spacetime symmetries of GR 
and its couplings to matter has been developed 
\cite{ak04}.
This framework provides conceptual insights
about the breaking of spacetime symmetries in various theories modifying GR
and remains the subject of ongoing investigation,
with numerous experiments performed to measure the coefficients
governing the symmetry breaking
\cite{tables}.
For ghost-free massive gravity,
it enables the derivation of constraints on matter couplings
from searches for Lorentz violation 
\cite{bbw19}.

In this work,
we investigate some specific implications 
of the breaking of spacetime symmetries in ghost-free massive gravity.
Our focus is primarily on properties related to the Lorentz transformations 
that emerge in approximately Minkowski spacetime,
although we expect many of the concepts
to apply in other background spacetimes as well.
These Lorentz transformations arise from a combination 
of local Lorentz and diffeomorphism transformations in curved spacetime
\cite{kl21}
and moreover are the focus of most experimental investigations
\cite{tables}, 
so they are of particular interest in the present context.
Here,
we explore two topics along these lines.
The first concerns the static extrema and saddle points of the potential
and their Lorentz structure.
We show the theory admits a variety of solutions
including ones that are Lorentz invariant 
and others that are Lorentz violating,
and we classify the patterns of symmetry breaking.
We determine the local, global, and absolute stability of the solutions,
verifying that an absolutely stable Lorentz-invariant extremum exists
for special potential parameters.

The second topic concerns the phenomenology
of flat non-Minkowski fiducial metrics,
$f_{\mu\nu}\not\propto \et_{\mu\nu}$.
In approximately Minkowski spacetime,
these can in principle generate observable signals 
of explicit Lorentz violation.
We study two scenarios of potential experimental interest,
gravitational waves and the limit of Newton gravity.
For gravitational waves representing excitations about Minkowski spacetime,
we demonstrate that the five modes of massive gravity
experience pentarefringence during propagation.
The modes are either superluminal or carry negative energies
in some observer frames,
a result typical of Lorentz-violating theories
\cite{kl01}. 
For Newton gravity,
we obtain the propagator and use it to determine
the gravitational potential energy between two point masses.
The modifications to the usual Yukawa form for the Fierz-Pauli case
include violations of both boost and rotation invariance.

The organization of this work is as follows.
In Sec. \ref{Setup},
we present essential preliminary material.
The action for massive gravity adopted here 
and the definitions of relevant variables 
are presented in Sec.\ \ref{Basics}.
A summary of the spacetime symmetries of the theory
is provided in Sec.\ \ref{Spacetime symmetries},
along with an outline of their implementation 
in approximately Minkowski spacetime.
In Sec.\ \ref{Matrix decomposition},
a decomposition of the key matrix variable 
convenient for calculational purposes is performed.

The investigation of extrema and saddle points of the action
is undertaken in Sec.\ \ref{Static extrema}.
We obtain the potential governing static solutions 
and solve the resulting equations of motion for flat fiducial metrics.
Three cases are distinguished,
treated in turn in Secs.\ \ref{Case A}, \ref{Case B}, and \ref{Case C}.
The static solutions are constructed 
and their Lorentz properties established.
In each case,
the issue of local stability, instability, or metastability 
is addressed using the technique of bordered hessians. 
The surface generated by the hamiltonian constraint 
and the positions of the solutions on its connected sheets 
are used to establish global and absolute stability properties.

In Sec.\ \ref{Linearized massive gravity},
the linearized limit of massive gravity is explored.
We investigate the Lorentz properties of solutions 
of the modified linearized Einstein equation 
that reduce to Minkowski spacetime
for vanishing fluctuations of the dynamical metric.
Gravitational waves are considered in Sec.\ \ref{Gravitational waves}.
We construct the eigenenergies and eigenmodes 
of the modified Einstein equation
and study their splitting for choices of the fiducial metric 
that differ from the Minkowski metric.
The limit of Newton gravity is examined in Sec.\ \ref{Propagator}.
The propagator for linearized massive gravity is obtained
and used to determine the gravitational potential 
energy between two stationary point masses.
An overall summary of the paper is provided in Sec.\ \ref{Summary},
while Appendix \ref{app:momentum-integrals}
describes technical details of certain integrals required in the text.
The conventions for metric, curvature, and other signs and factors
are those of Ref.\ \cite{ak04}, Appendix A.

\section{Setup}
\label{Setup}

In this section,
we discuss the form of the action for massive gravity 
used in the present work.
We summarize key aspects of spacetime symmetries and their violations,
and we provide a matrix decomposition
for the dynamical variable of central interest in the analyses to follow.

\subsection{Basics}
\label{Basics}

Consider the action for ghost-free massive gravity in the form 
\cite{hr11}
\beq
S = \frac 1 {2\ka} \int d^4x \sqrt{-g}
\Big( R - 2m^2 \sum_{n=0}^4 \be_n e_n(\mathbb{X}) \Big) 
+ S_{\rm m},
\label{S}
\eeq
where $\ka = 8 \pi G$.
The first term is the usual Einstein-Hilbert action
written in terms of the dynamical metric $g_{\mu\nu}$
and the Riemann scalar curvature $R$.
The second term can be understood as a nonderivative scalar potential 
for the dynamical metric, 
which depends on the mass scale $m$
and on five dimensionless parameters $\be_n$, $n= 0,\ldots, 4$.
Only two combinations of the latter have independent physical meaning
\cite{hr11}.
To ensure that the matrix of Poisson brackets 
of second class constraints is invertible,
the parameters satisfy the condition \cite{sa14}
$\be_1 + 2\be_2 + \be_3 \neq 0$.

The potential in the action \rf{S}
also involves a nondynamical fiducial metric $f_{\mu\nu}$, 
which can be chosen as desired
but is often taken to be the Minkowski metric $\et_{\mu\nu}$.
The fiducial metric appears in the action in the combination
$\mathbb{X}^\mu{}_\nu = (\sqrt{g^{-1}f}~)^\mu{}_\nu$,
where the matrix $g^{-1}f$ is assumed to have only positive eigenvalues
so that its square root is well defined.
The five invariant polynomials $e_n(\mathbb{X})$
involve traces of powers of the argument of $\mathbb{X}^\mu{}_\nu$.
They are defined by $e_0(\mathbb{X})=1$ and the recursive relation 
\cite{hr11}
\beq
e_n(\mathbb{X}) = 
-\frac{1}{n}\sum_{k=1}^n (-1)^k[\mathbb{X}^k] e_{n-k}(\mathbb{X}),
\label{recursive-e_n}
\eeq
where $[\cdot]$ indicates a trace,
$[\mathbb{Z}]=\mathbb{Z}^\mu{}_\mu$.
It follows that
\bea
e_0(\mathbb{X}) &=& 1,
\nn\\
e_1(\mathbb{X}) &=& [\mathbb{X}] \equiv \mbox{tr}~\mathbb{X} ,
\nn\\
e_2(\mathbb{X}) &=& \tfrac{1}{2}\left([\mathbb{X}]^2 - [\mathbb{X}^2]\right),
\nn\\
e_3(\mathbb{X}) &=& 
\tfrac{1}{6}\left([\mathbb{X}]^3 - 3[\mathbb{X}][\mathbb{X}^2]
+ 2[\mathbb{X}^3]\right),
\nn\\
e_4(\mathbb{X})\hskip-2pt &=& \hskip-2pt 
\tfrac{1}{24}\left([\mathbb{X}]^4 \hskip-2pt - 6[\mathbb{X}]^2[\mathbb{X}^2]
\hskip-2pt + 3[\mathbb{X}^2]^2 \hskip-2pt 
+ 8[\mathbb{X}][\mathbb{X}^3] \hskip-2pt - 6[\mathbb{X}^4]\right)
\nn\\
&\equiv&\mbox{det}\,\mathbb{X}.
\label{mgpotfun}
\eea
In four spacetime dimensions,
$e_n(\mathbb{X}) \equiv 0$ for $n\ge5$.
The term with $n=4$ is nondynamical and hence can be omitted
from the action \rf{S}. 
The action can be extended 
by the addition of a matter term $S_{\rm m}$ 
to describe the coupling of matter as desired.

The analyses in the present work take advantage 
of a duality property of the scalar potential
\cite{hr12},
\beq
\sqrt{-g}\,e_n(\mathbb{X})=\sqrt{-f}\,e_{4-n}(\mathbb{Y}),
\label{duality}
\eeq
where
\beq
\mathbb{Y}^\mu{}_\nu = \left(\mathbb{X}^{-1}\right)^\mu{}_\nu
= (\sqrt{f^{-1}g}~)^\mu{}_\nu .
\label{defY}
\eeq
For our purposes,
it is advantageous to work with the representation of the scalar potential 
in terms of the matrix $\mathbb{Y}^\mu{}_\nu$ 
rather than $\mathbb{X}^\mu{}_\nu$.
This replaces the determinant $\sqrt{-g}$ of the dynamical metric
with the determinant $\sqrt{-f}$ of the nondynamical fiducial metric
and thereby simplifies the variational analysis.
The second term in the action \rf{S} then takes the form 
$-(m^2/\ka)\int d^4x \sqrt{-f}\,\mathcal{V}$,
where the potential $\mathcal{V}$ is given by 
\beq
\mathcal{V}(\mathbb{Y}) = \sum_{n=0}^4\ol{\be}_n e_n(\mathbb{Y}),
\label{V}
\eeq
with
\beq
\ol\be_n = \be_{4-n} .
\eeq
The parameters $\ol\be_n$ are related to the parameters $\al_n$ 
defined in Part II of the review in Ref.\ \cite{dr14} by 
\beq
\left(
\begin{matrix}
\ol\be_0\\
\ol\be_1\\
\ol\be_2\\
\ol\be_3\\
\ol\be_4
\end{matrix}
\right)
=
\left(
\begin{matrix}
0 & 0 & 0 & 0 & 1 \\
 0 & 0 & 0 & -1 & -4 \\
 0 & 0 & 1 & 3 & 6 \\
 0 & -1 & -2 & -3 & -4 \\
 1 & 1 & 1 & 1 & 1
\end{matrix}
\right)
\left(
\begin{matrix}
\al_0\\
\al_1\\
\al_2\\
\al_3\\
\al_4
\end{matrix}
\right),
\eeq
with inverse
\beq
\left(
\begin{matrix}
\al_0\\
\al_1\\
\al_2\\
\al_3\\
\al_4
\end{matrix}
\right)=
\left(
\begin{matrix}
 1 & 1 & 1 & 1 & 1 \\
 -4 & -3 & -2 & -1 & 0 \\
 6 & 3 & 1 & 0 & 0 \\
 -4 & -1 & 0 & 0 & 0 \\
 1 & 0 & 0 & 0 & 0
\end{matrix}
\right)
\left(
\begin{matrix}
\ol\be_0\\
\ol\be_1\\
\ol\be_2\\
\ol\be_3\\
\ol\be_4
\end{matrix}
\right).
\eeq
The parameter $\al_0$ corresponds to the cosmological constant,
$\al_1$ to a tadpole contribution,
$\al_2$ to a mass term generalizing the Fierz-Pauli action,
while $\al_3$ and $\al_4$ correspond to higher-order interactions.

\subsection{Spacetime symmetries}
\label{Spacetime symmetries}

The spacetime symmetries of the action \rf{S} are keys to its physical content.
This subsection provides a brief summary of 
some features of particular interest in what follows.  
A recent discussion with more details and in a broader context 
can be found in Ref.\ \cite{kl21}.

In considering spacetime symmetries of a theory,
particularly one containing nondynamical backgrounds
like the fiducial metric $f_{\mu\nu}$ in the action \rf{S},
it is useful to define two classes of transformations
\cite{ck97,ak04}.
Observer transformations change the observer frame,
and hence they amount to coordinate choices that leave unaffected the physics. 
Geometrically,
they act the atlas of the spacetime manifold.
Particle transformations change dynamical particles and fields,
leaving invariant nondynamical quantities 
and thus modifying their couplings.
Geometrically,
particle transformations act on the spacetime manifold 
and its tangent and cotangent bundles. 
A physical symmetry under a particle transformation may therefore be violated
by the presence of a nondynamical quantity
even if the theory is invariant under the corresponding observer transformation.
Note that the two classes of transformations expressed in a coordinate basis
are mathematically similar when nondynamical quantities are absent
and are then sometimes called passive and active,
but this similarity fails in the general scenario.

{\it General coordinate transformations}
are prime examples of observer transformations,
implementing smooth coordinate changes 
and leaving invariant the action \rf{S}.
For explicit calculations,
a particular set of coordinates is often selected,
corresponding to a convenient choice of observer frame.
The fiducial metric $f_{\mu\nu}$ behaves as a 2-form 
under general coordinate transformations,
so a suitable choice of coordinates can bring it to a convenient form. 
The dynamical metric $g_{\mu\nu}$ is also a 2-form 
under general coordinate transformations,
so the choice of coordinates affects its explicit form as well. 

{\it Local Lorentz transformations} are particle transformations
that act on the tangent space at each point on the spacetime manifold.
They thus change quantities defined in a local frame
while leaving unaffected ones defined in a spacetime frame.
We label spacetime coordinates by Greek indices $\mu$, $\nu$, $\ldots$
and local coordinates by Latin indices $a$, $b$, $\ldots$.
The dynamical metric $g_{\mu\nu}$ in a local frame
can be related to the nondynamical Minkowski metric $\et_{ab}$
through the dynamical vierbein $\vb\mu a$ according to
$g_{\mu\nu}=\vb\mu a\vb\nu b\et_{ab}$.
A local Lorentz transformation at point $x$
described by the matrix components $\La^a_{\pt{a}b}(x)$ 
acts on $\vb\mu a$, $g_{\mu\nu}$, and $f_{\mu\nu}$ as
\bea
e_\mu^{\pt{\mu}a}
&\to& 
\La^a_{\pt{a}b} e_\mu^{\pt{\mu}b},
\nn\\
g_{\mu\nu}
&\to& g_{\mu\nu},
\nn\\
f_{\mu\nu}
&\to& f_{\mu\nu}.
\label{LLT}
\eea
Combinations of objects such as the matrix $\mathbb{Y}^\mu{}_\nu$
inherit the corresponding transformation properties.
Since the action \rf{S} is specified in terms of objects
defined in a spacetime frame,
it is invariant under local Lorentz transformations.

{\it Diffeomorphisms} are particle transformations
consisting of smooth maps of the spacetime manifold
into itself and hence embody the notion of local translations,
with a spacetime point $x^\mu$ mapped to another point
according to $x^\mu \to x^{\prime\mu}=x^\mu+\xi^\mu (x)$
when expressed in a fixed coordinate system.
Under a diffeomorphism,
dynamical quantities transform as the induced pushforward or pullback.
For infinitesimal diffeomorphisms,
dynamical quantities transform via the Lie derivative,
while nondynamical quantities remain unaffected.
For example,
under an infinitesimal diffeomorphism
the vierbein, dynamical metric, and fiducial metric transform as
\bea
e_\mu{}^a 
&\to&
\vb \mu a - \vb \rh a \prt_\mu \xi^\rh -\xi^\la \prt_\la \vb \mu a,
\nn\\
g_{\mu\nu}
&\to&
g_{\mu\nu}-g_{\rh\nu} \prt_\mu \xi^\rh -g_{\mu\si} \prt_\nu \xi^\si
-\xi^\la \prt_\la g_{\mu\nu},
\nn\\ 
f_{\mu\nu}
&\to& 
f_{\mu\nu}.
\label{DT}
\eea
As a result,
diffeomorphism invariance is broken 
by all the terms in the potential for the action \rf{S}
except the term proportional to $\be_0$,
which acts as a cosmological constant.

{\it Manifold Lorentz transformations}
are particle transformations that act 
both on spacetime points and on local frames 
as combinations of special diffeomorphisms and local Lorentz transformations.
They are of particular interest in the present context
because they are the analogues in approximately Minkowski spacetime
of global Lorentz transformations in Minkowski spacetime
\cite{kl21}.
Given a fixed $\La$ in the Lorentz group,
the corresponding manifold Lorentz transformation
consists of the special diffeomorphism
$x^\mu \to x^{\prime\mu}=\La^\mu_{\pt{\mu}\nu} x^\nu$
mapping each spacetime point $x^\mu$ to another point $x^{\prime\mu}$ 
via the matrix $\La^\mu_{\pt{\mu}\nu}$ in a fixed coordinate system,
along with a special local Lorentz transformation
such that the vierbein and metrics transform at each $x$ as
\bea
\vb \mu a  &\to& 
(\La^{-1})^\rh_{\pt{\rh}\mu} \La^a_{\pt{a}b} \vb \rh b,
\nn\\
g_{\mu\nu} &\to& 
(\La^{-1})^\rh_{\pt{\rh}\mu} (\La^{-1})^\si_{\pt{\si}\nu} g_{\rh\si},
\nn\\
f_{\mu\nu} &\to& f_{\mu\nu}.
\label{mlt}
\eea
In part of this work,
we investigate features of massive gravity
in approximately Minkowski spacetime,
where the dynamical metric $g_{\mu\nu}$
contains only small fluctuations 
away from the Minkowski metric $\et_{\mu\nu}$
and the manifold Lorentz transformations
reduce to the usual notion of Lorentz transformations 
in approximately Minkowski spacetime.
Within this scenario,
the action \rf{S} is invariant under Lorentz transformations
whenever the fiducial metric is constant 
and proportional to the Minkowski metric,
$f_{\mu\nu}\propto\et_{\mu\nu}$,
because the transformation law \rf{mlt} for $f_{\mu\nu}$
then coincides with the standard Lorentz transformation
under which $\et_{\mu\nu}$ is invariant.
However,
for other fiducial metrics $f_{\mu\nu}\not\propto\et_{\mu\nu}$,
the transformation law \rf{mlt} for $f_{\mu\nu}$
lacks the usual action of $\La^\mu_{\pt{\mu}\nu}$ 
and so the action \rf{S} violates Lorentz invariance.
We thus see that the diffeomorphism violation arising from the mass term
transcribes to Lorentz violation in approximately Minkowski spacetime 
except for the special choice $f_{\mu\nu}\propto\et_{\mu\nu}$.
Note also that violations of rotation symmetry are embedded 
in Lorentz violation 
because rotations form a subgroup of the Lorentz group.

{\it CPT transformations} can be understood 
in Minkowski spacetime as the product 
of charge conjugation C, parity inversion P, and time reversal T.
They are closely linked to global Lorentz transformations
in Minkowski spacetime,
with the link formally being established via the CPT theorem 
\cite{cpt}.
In curved spacetime,
CPT is challenging to define but a practical implementation exists 
\cite{ak04}.
Under this implementation,
the action \rf{S} is CPT invariant even for nontrivial curvature.
In approximately Minkowski spacetime,
a CPT transformation paralleling the usual one can be constructed.
CPT invariance is then a feature of local realistic theories 
containing backgrounds carrying an even number of spacetime indices,
which includes the fiducial metric $f_{\mu\nu}$.
The action \rf{S} therefore exhibits CPT invariance
in approximately Minkowski spacetime as well.

We remark in passing that the implementation
of the above spacetime symmetries 
in alternative formulations of massive gravity
may require separate consideration.
For example,
the alternative vierbein formulation 
\cite{hir12}
using the dynamical vierbein $\vb \mu a$
and a nondynamical fiducial vierbein $f_\mu{}^a$
explicitly violates both local Lorentz and diffeomorphism invariances
because $f_\mu{}^a$ fails to transform conventionally
\cite{kl21}.
However,
if the vierbeins satisfy the condition 
$e^\mu{}_a f_\mu{}^c \et_{cb} = \et_{ac} f_\mu{}^c e^\mu{}_b$,
then this alternative formulation is equivalent to the action \rf{S}
\cite{hir12,dmz13}
and so local Lorentz invariance is preserved.
As another example,
bimetric massive gravity
\cite{hr12}
involves two dynamical metrics $g_{\mu\nu}$, $f_{\mu\nu}$.
Their background values emerging from extremizing the bimetric action 
therefore must solve the equations of motion,
which implies any Lorentz breaking is spontaneous
and accompanied by massless fluctuations
\cite{bk05},
which are Nambu-Goldstone modes
\cite{ng60}.
Techniques are available for handling 
the resulting phenomenological complications 
\cite{bkt06},
and many experiments have sought the corresponding effects
\cite{tables}.
Investigating the implications for bimetric massive gravity
of these results and of the methods discussed here
would be of definite interest but lies outside our present scope.
Note that in contrast no fluctuations are associated with 
the nondynamical fiducial metric $f_{\mu\nu}$ in the action \rf{S},
where the Lorentz breaking is explicit.
The phenomenology of explicit breaking without fluctuations 
can be explored in gravitational effective field theory
\cite{kl21two}. 

Given that manifold Lorentz symmetry is generically violated 
in the action \rf{S},
it is of interest to determine the pattern of the symmetry breaking 
in any given scenario.
As an illustration of some relevant ideas consider
the analysis in Sec.\ \ref{Static extrema} below 
of the extrema and saddle points of the potential \rf{V},
which is a quartic in the matrix variable $\mathbb{Y}^\mu{}_\nu$.
The equation determining the extrema and saddle points is therefore a cubic,
with three independent solutions.
Since $\mathbb{Y}^\mu{}_\nu$ has at most four different eigenvalues,
it follows that at least two of them must be degenerate 
and so at most five of the six Lorentz generators can break.
If three eigenvalues are degenerate,
then three Lorentz generators are broken,
while if two pairs of eigenvalues are degenerate 
then four Lorentz generators break.
The above line of reasoning 
reveals that the basic structure of the potential \rf{V}
excludes one, two, or six broken Lorentz generators.
As we show in Sec.\ \ref{Static extrema},
the cubic governing the extrema and saddle points of the potential \rf{V}
has degenerate roots, 
and so in fact the only solutions either are Lorentz invariant
or have four broken Lorentz generators.
In the latter case,
the pattern of symmetry breaking is
SO(1,3) $\to$ SO(1,1) $\times$ SO(2).

This pattern differs from ones 
known in other Lorentz-violating models of gravity.
Consider,
for example,
the cardinal model
\cite{cardinal},
which is also an extension of GR containing a nonlinear potential.
It is constructed starting in Minkowski spacetime 
with a symmetric 2-tensor that undergoes spontaneous Lorentz violation
and requiring self-consistent coupling to the energy-momentum tensor.
In the Lorentz-invariant case,
this bootstrap procedure is known to generate GR from a massless spin-2 field
\cite{bootstrap}. 
In the cardinal model,
a unique combination satisfies the integrability condition
for self consistency at each order in the field fluctuations
\cite{ms20}.
The potential functions for the cardinal model,
defined by Eq.\ (134) of Ref.\ \cite{cardinal},
match the polynomials \rf{mgpotfun} for ghost-free gravity
but serve as input for the differential equations 
satisfied by the bootstrap potentials
rather than being combined to eliminate the ghost.
The cardinal model is thus a bootstrap theory like GR
but generically contains a ghost,
while the action \rf{S} for massive gravity is ghost free
but generically cannot be obtained via a bootstrap. 
Known patterns of Lorentz breaking for the cardinal potential 
exclude situations with one or two broken Lorentz generators,
as before.
However,
they include ones with three, five, and six broken generators
\cite{ctw09},
which cannot occur in ghost-free massive gravity
as outlined above.

\subsection{Matrix decomposition}
\label{Matrix decomposition}

The analysis of extrema and saddle points in Sec.\ \ref{Static extrema}
is performed using the matrix variable $\mathbb{Y}^\mu{}_\nu$
defined in Eq.\ \rf{defY},
with the special choice of fiducial metric $f_{\mu\nu}=\et_{\mu\nu}$.
This subsection provides a decomposition of $\mathbb{Y}^\mu{}_\nu$
in terms of variables convenient for the subsequent derivations.

The square of $\mathbb{Y}^\mu{}_\nu$
can be written using the Arnowitt-Deser-Misner decomposition
\cite{adm59},
\beq
\bigl(\mathbb{Y}^2\bigr)^\mu{}_\nu=
\bigl(\et^{-1}g\bigr)^\mu{}_\nu=
\left(
\begin{matrix}
N^2-N_i\ga^{ij}N_j \ & -N_j\\
N_j & \ga_{ij}
\end{matrix}
\right),
\label{ADM}
\eeq
where $\ga_{ij}=g_{ij}$ is the spacelike part of the dynamical metric
with inverse $\ga^{ij}$,
$N_i=g_{0i}$ is the shift variable,
and $N=(-g^{00})^{-1/2}$ is the lapse. 
In the GR action,
the shift and the lapse appear linearly
and multiply first class constraints.
In the case of massive gravity,
however,
the potential term in the action destroys linearity,
and the variables $N_\mu = (N,N_i)$ acquire equations of motion
determining them in terms of the dynamical fields.
This leaves $10-4=6$ propagating modes,
including the Boulware-Deser ghost.

To eliminate the ghost from the spectrum,
the equations of motion for $N_\mu$ 
must involve only three of the four degrees of freedom.
This means that the equations of motion 
depend on only three combinations $n_i$ of the four variables $N_\mu$,
along with the metric variables $\ga_{ij}$.
It is then natural to perform a change of variables
\beq
\{N_i,N,\ga_{ij}\} \to \{n_i,N,\ga_{ij}\}
\label{N->n}
\eeq
that eliminates the $N_i$ in favor of the $n_i$.
The $n_i$ are auxiliary fields fixed by their own equations of motion.
The lapse $N$ does not appear in its own equation of motion
and hence acts as a Lagrange multiplier multiplying a constraint.
This constraint eliminates the Boulware-Deser ghost.
It follows that the potential must be linear in $N$
after performing the change of variables \rf{N->n}.
Given the form of the decomposition \rf{ADM},
the transformation of $N_i$ must therefore be linear in $N$,
\beq
N_i = \bigl( \de_i{}^j + N D_i{}^j \bigr) n_j .
\label{N->n_expl}
\eeq
The matrix $D_i{}^j$ is determined by the requirement 
that the action be linear in $N$.

To implement this line of reasoning explicitly for $\mathbb{Y}^\mu{}_\nu$,
we take
\beq
\mathbb{Y}^\mu{}_\nu =
\bigl(\sqrt{\et^{-1}g}\bigr)^\mu{}_\nu =
\mathbb{A}^\mu{}_\nu + N \mathbb{B}^\mu{}_\nu .
\label{al-be}
\eeq 
Squaring then gives
\beq
\bigl(\mathbb{Y}^2\bigr)^\mu{}_\nu =
\bigl(\mathbb{A}^2\bigr)^\mu{}_\nu 
+ N (\mathbb{A}.\mathbb{B} + \mathbb{B}.\mathbb{A})^\mu{}_\nu
+ N^2 \bigl(\mathbb{B}^2\bigr)^\mu{}_\nu .
\label{al-be-2}
\eeq 
We can compare this result with the decomposition \rf{ADM},
using the expression \rf{N->n_expl}.
It follows that 
\begin{align}
\mathbb{A}^2 &=
\left(
\begin{matrix}
-n^T\ga^{-1}n \ & -n^T\\
n & \ga
\end{matrix}
\right),
\nn\\
\mathbb{B}^2 &=
\bigl( 1 - n^T D^T \ga^{-1} D n\bigr)
\left(
\begin{matrix}
1 \ & 0\\
0 & 0
\end{matrix}
\right),
\nn\\
\mathbb{A}.\mathbb{B} + \mathbb{B}.\mathbb{A} &=
\left(
\begin{matrix}
-n^T D^T\ga^{-1} n - n^T\ga^{-1} D n \ & -n^T D^T\\
D n & 0
\end{matrix}
\right) .
\label{al.be-be.al}
\end{align}
The first two of these identities determine
the matrices $\mathbb{A}$ and $\mathbb{B}$ upon taking matrix square roots,
which are uniquely defined if $\mathbb{A}^2$ and $\mathbb{B}^2$ 
are diagonalizable with nonnegative eigenvalues.
This is indeed the case for sufficiently small values of $n_i$.
We find
\beq
\mathbb{B} =
\sqrt{1 - n^T D^T \ga^{-1} D n}
\begin{pmatrix}
1 \ & 0\\
0 & 0
\end{pmatrix} .
\label{be}
\eeq
Choosing coordinates such that $\ga = \mathbb{1}$,
we also obtain
\beq
\mathbb{A} =
\frac{1}{\nnn}
\begin{pmatrix}
-n^T n\ & -n^T\\
n\ & \mathbb{1}\nnn 
+ n n^T\,\dfrac{1 - \nnn}{n^T n}
\end{pmatrix} ,
\label{al}
\eeq
where $\nnn \equiv {\sqrt{1 - n^T n}}$.
Note that this expression is independent of the matrix $D$.

The results \rf{be} and \rf{al} for the matrices $\mathbb{A}$ and $\mathbb{B}$
can now be used to find the explicit form of $\mathbb{Y}^0{}_\nu$.
This gives
\bea
\mathbb{Y}^0{}_0 &=& \mathbb{A}^0{}_0 + N\,\mathbb{B}^0{}_0
= \frac{-n^Tn}{\sqrt{1 - n^Tn}} + N \nnn \equiv \wtnn,
\nn\\
\mathbb{Y}^0{}_i &=& \mathbb{A}^0{}_i + N\,\mathbb{B}^0{}_i
= \frac{-n_i}{\sqrt{1 - n^Tn}} \equiv -\wtn_i,
\label{tilde-ni}
\eea
where we introduced the convenient variables $\wtnn$ and $\wtn_i$.
The inverse relations are
\beq
n_i = \frac{\wtn_i}{\sqrt{1 + \wtn^T\wtn}},
\quad
N = \sqrt{1 + \wtn^T\wtn} \wtnn + \wtn^T\wtn,
\eeq
from which the partial derivative with respect to $N$
is found to be
\beq
\frac{\partial}{\partial N} = 
\frac{\partial\tilde N}{\partial N}\frac{\partial}{\partial \wtnn}
+ \frac{\partial\tilde n_i}{\partial N}\frac{\partial}{\partial \wtn_i}
= \nnn \frac{\partial}{\partial \wtnn}.
\label{dN}
\eeq
The partial derivative with respect to $n_i$ is a linear combination
of the partial derivatives with respect to $\wtn_i$ and $\wtnn$.
Note that the hamiltonian constraint is obtained by taking 
the partial derivative with respect to the lapse $N$,
so the result \rf{dN} implies it can alternatively be obtained 
by taking the partial derivative with respect to
$\wtnn = \mathbb{Y}^0{}_0$.

\section{Static solutions}
\label{Static extrema}

Among the solutions obtained by varying the action \rf{S}
are static ones with vanishing curvatures
for both the metrics $g_{\mu\nu}$ and $f_{\mu\nu}$,
which can be interpreted as flat vacuum spacetimes.
In this section,
these solutions are classified and constructed.
We take advantage of general coordinate invariance
to choose a special observer frame in which
the fiducial metric takes the form 
of the Minkowski metric $f_{\mu\nu}=\et_{\mu\nu}$,
and we determine the corresponding solutions 
to the static equations of motion 
for the matrix variable $\mathbb{Y}^\mu{}_\nu$.
The explicit form of solutions for any other fiducial metric 
$f_{\mu\nu}\not\propto\et_{\mu\nu}$
can then be obtained via a suitable general coordinate transformation.

The extrema and saddle points of interest are solutions
of the equations of motion obtained by varying the potential \rf{V}.
Since the term with parameter $\ol\be_0$ is constant,
it can be set to zero without loss of generality in the analysis.
It therefore suffices to study the equations of motion 
obtained from the potential
\beq
\mathcal{U}(\mathbb{Y})	= \sum_{i=1}^4\ol{\be}_i e_i(\mathbb{Y}).
\label{U}
\eeq

For the analysis,
it is convenient to parametrize $\mathbb{Y}^\mu{}_\nu$ as
\beq
\mathbb{Y}^\mu{}_\nu =
\begin{pmatrix}
\wtnn & -\wtn_i \\
\wtn_j & k_{ij} 
\end{pmatrix}.
\label{potential}
\eeq
The solutions for the variables $\wtn_i$ can be obtained directly 
from their equations of motion,
and we find
\beq
\wtn_i = 0 .
\eeq

To make further progress,
we diagonalize the spacelike part $k_{ij}$ of $\mathbb{Y}^\mu{}_\nu$ 
by applying an othogonal transformation
$\mathbb{Y} \to \mathcal{O}\mathbb{Y}\mathcal{O}^T$,
which amounts to a field redefinition
and thus leaves the physics unchanged.
This brings $\mathbb{Y}^\mu{}_\nu$ to the form
\beq
\mathbb{Y}^\mu{}_\nu =
\begin{pmatrix}
\wtnn & 0 & 0 & 0 \\
0 & \la_1 & 0 & 0 \\
0 & 0 & \la_2 & 0 \\
0 & 0 & 0 & \la_3 
\end{pmatrix}.
\label{ygauge}
\eeq
The potential \rf{U} then becomes
\bea
\mathcal{U}(\mathbb{Y}) &=& 
\ol{\be}_1(\la_1 + \la_2 + \la_3) 
+ \ol{\be}_2(\la_1 \la_2 + \la_2\la_3
+ \la_3\la_1) 
\nn\\
&&
+ \ol{\be}_3\la_1\la_2\la_3 
\nn\\
&&
+ \wtnn\bigl[\ol{\be}_1 
+ \ol{\be}_2(\la_1 + \la_2 + \la_3) 
\nn\\
&&
\hskip 10pt
+ \ol{\be}_3(\la_1 \la_2 + \la_2\la_3 + \la_3\la_1) 
+ \ol{\be}_4\la_1\la_2\la_3\bigr],
\quad
\label{potential2}
\eea
and it depends on the four field variables 
$\wtnn$, $\la_1$, $\la_2$, $\la_3$.

The equation of motion for $\wtnn$ yields the hamiltonian constraint,
\bea
&&
\ol{\be}_1 + \ol{\be}_2(\la_1 + \la_2 + \la_3)
+ \ol{\be}_3(\la_1 \la_2 + \la_2\la_3 + \la_3\la_1) 
\nn\\
&&
\hskip30pt
+ \ol{\be}_4\la_1\la_2\la_3 = 0.
\label{tilde-N-eom}
\eea
The equations of motion for the remaining three variables
$\la_1$, $\la_2$ and $\la_3$ are 
\bea
&&
\ol{\be}_1 
+ \ol{\be}_2(\wtnn + \la_2 + \la_3) 
+ \ol{\be}_3\bigl(\wtnn(\la_2 + \la_3) 
+ \la_2\la_3\bigr) 
\nn\\
&&
\hskip30pt
+ \ol{\be}_4\wtnn\la_2\la_3 = 0 ,
\nn\\
&&
\ol{\be}_1 
+ \ol{\be}_2(\wtnn + \la_1 + \la_3) 
+ \ol{\be}_3\bigl(\wtnn(\la_1 + \la_3) 
+ \la_1\la_3\bigr) 
\nn\\
&&
\hskip30pt
+ \ol{\be}_4\wtnn\la_1\la_3 = 0 ,
\nn\\
&&
\ol{\be}_1 
+ \ol{\be}_2(\wtnn + \la_1 + \la_2) 
+ \ol{\be}_3\bigl(\wtnn(\la_1 + \la_2) 
+ \la_1\la_2\bigr) 
\nn\\
&&
\hskip30pt
+ \ol{\be}_4\wtnn\la_1\la_2 = 0 .
\label{la_3-eom}
\eea
The parameter $\ol{\be_1}$ can be eliminated from these three equations 
by working instead with their differences.
For example,
subtracting the second equation from the first yields 
\beq
(\la_2 - \la_1)
\bigl[\ol{\be}_2 + \ol{\be}_3(\wtnn + \la_3)
+ \ol{\be}_4 \wtnn\la_3\bigr] = 0.
\label{la_2-la_1-eom}
\eeq
Similarly,
we find 
\beq
(\la_3 - \la_1)
\bigl[\ol{\be}_2 + \ol{\be}_3(\wtnn + \la_2)
+ \ol{\be}_4 \wtnn\la_2\bigr] = 0,
\label{la_3-la_1-eom}
\eeq
and
\beq
(\la_3 - \la_2)
\bigl[\ol{\be}_2 + \ol{\be}_3(\wtnn + \la_1)
+ \ol{\be}_4 \wtnn\la_1\bigr] = 0.
\label{la_3-la_2-eom}
\eeq
In what follows,
we solve the system of these equations 
and the hamiltonian constraint \rf{tilde-N-eom}
for each of three cases in turn:
case A with $\ol{\be}_4 \ne 0$,
case B with $\ol{\be}_4 = 0$, $\ol{\be}_3 \ne 0$,
and case C with 
$\ol{\be}_4 = \ol{\be}_3 = 0$, $\ol{\be}_2 \ne 0$.
This establishes the complete set of desired static extrema
and saddle points of the action \rf{S}.

\subsection{Case A: $\ol{\be}_4 \ne 0$}
\label{Case A}

Consider first case A with $\ol{\be}_4 \ne 0$.
We obtain here the solutions 
determine their local stability,
and investigate global stability
for the subset of locally stable configurations.

\subsubsection{Static solutions}

Inspection reveals that one class of solutions 
of Eqs.\ \rf{la_2-la_1-eom}--\rf{la_3-la_2-eom}
is obtained by taking $\la_1 = \la_2 = \la_3$. 
Substitution into Eq.\ \rf{tilde-N-eom} yields the cubic equation
\beq
\ol{\be}_1 + 3\ol{\be}_2\la_1 
+ 3\ol{\be}_3\la_1^2 + \ol{\be}_4\la_1^3 = 0.
\label{cubic-la_1-eq}
\eeq
For the case $\ol{\be}_4 \ne 0$,
this yields either one or three real solutions for $\la_1$.
In terms of the discriminant
\beq
D = 4\ol{\be}_4\ol{\be}_2^3 
- 3\ol{\be}_3^2\ol{\be}_2^2 
- 6\ol{\be}_1\ol{\be}_2\ol{\be}_3\ol{\be}_4 
+ 4 \ol{\be}_1\ol{\be}_3^3 + \ol{\be}_1^2\ol{\be}_4^2,
\label{D}
\eeq
the cubic \rf{cubic-la_1-eq} 
has three distinct real roots iff $D < 0$,
two coincident real roots if $D = 0$,
and one real root iff $D>0$. 
From the first expression in Eq.\ \rf{la_3-eom} it follows that
\beq
\wtnn = 
-\frac{\ol{\be}_1 + 2\ol{\be}_2\la_1 + \ol{\be}_3\la_1^2}
{\ol{\be}_2 + 2\ol{\be}_3\la_1 + \ol{\be}_4\la_1^2} 
= \la_1.
\eeq
The second equality is obtained 
by substitution of the solution for $\ol{\be}_1$
obtained from Eq.\ \rf{cubic-la_1-eq}.
We conclude that these solutions obey 
\beq
\wtnn = \la_1 = \la_2 = \la_3,
\label{caseA1}
\eeq 
with all four variables given by a single root of the cubic \rf{cubic-la_1-eq}.
The matrix $\mathbb{Y}^\mu{}_\nu$ is therefore proportional to the identity,
which implies this class of solutions is manifold Lorentz invariant.
The existence of three real roots ensures that three distinct solutions occur. 

It can be verified from Eqs.\ \rf{la_2-la_1-eom}--\rf{la_3-la_2-eom}
that the variables $\la_1$, $\la_2$, $\la_3$ cannot all be different.
However,
a second class of solutions can be obtained by setting 
any two of the $\la_i$ equal,
while keeping the third distinct.
Suppose for definiteness that $\la_1 = \la_2 \ne \la_3$.
It then follows from Eq.\ \rf{la_3-la_1-eom} that
\beq
\wtnn = 
-\frac{\ol{\be}_2 + \ol{\be}_3\la_1}{\ol{\be}_3 
+ \ol{\be}_4\la_1},
\label{sol-tilde-N}
\eeq
while Eq.\ \rf{tilde-N-eom} yields 
\beq
\la_3 = 
-\frac{\ol{\be}_1 + 2\ol{\be}_2\la_1 
+ \ol{\be}_3\la_1^2}{\ol{\be}_2 
+ 2\ol{\be}_3\la_1 + \ol{\be}_4\la_1^2} .
\label{sol-la_3}
\eeq
From Eqs.\ \rf{tilde-N-eom} and \rf{la_3-eom},
we find
\beq
(\la_3 - \wtnn)
\bigl[\ol{\be}_2 + 2\ol{\be}_3 \la_1
+ \ol{\be}_4 \la_1^2\bigr] = 0 .
\label{la_3-tilde_N-eq}
\eeq
Taking the second factor in this equation to vanish
leads to a divergent expression 
on the right-hand side of Eq.\ \rf{sol-la_3}, 
so $\la_3 = \wtnn$ is required.
Combining this result with Eqs.\ \rf{sol-tilde-N} and \rf{sol-la_3}
then yields the identity
\beq
\ol{\be}_1\ol{\be}_3 - \ol{\be}_2^2 
+ (\ol{\be}_1\ol{\be}_4 - \ol{\be}_2\ol{\be}_3)\la_1 
+ (\ol{\be}_2\ol{\be}_4 - \ol{\be}_3^2)\la_1^2 = 0 ,
\label{quadratic-eq}
\eeq
with the solutions
\beq
\la_1 = \la_2 = 
\frac{\ol{\be}_2\ol{\be}_3 
- \ol{\be}_1\ol{\be}_4 \pm \sqrt{D}}{2(\ol{\be}_2\ol{\be}_4 
- \ol{\be}_3^2)} .
\label{la_LV}
\eeq
Note the appearance of the discriminant \rf{D},
with the solutions being real iff $D > 0$.
Using Eq.\ \rf{sol-tilde-N} then reveals that
\beq
\la_3 = \wtnn =
-\frac{\ol{\be}_2 
+ \ol{\be}_3\la_1}{\ol{\be}_3 + \ol{\be}_4\la_1} 
= \frac{\ol{\be}_2\ol{\be}_3 
- \ol{\be}_1\ol{\be}_4 \mp \sqrt{D}}{2(\ol{\be}_2\ol{\be}_4 
- \ol{\be}_3^2)}.
\eeq
The second class of solutions therefore obeys
\beq
\wtnn = \la_3 \neq \la_1 = \la_2,
\label{caseA2}
\eeq 
with the two subsets of equal variables 
specified as the two roots of the quadratic \rf{quadratic-eq}.
Since the matrix $\mathbb{Y}^\mu{}_\nu$ differs from the identity,
this class of solutions violates manifold Lorentz invariance.
The pattern of symmetry breaking is
SO(1,3) $\to$ SO(1,1) $\times$ SO(2).
Note that we can obtain two more pairs of analogous solutions 
by interchanging the role of $\la_3$ with $\la_1$ and $\la_2$ in turn.
Note also that the solutions are obtained
assuming the form \rf{ygauge} for $\mathbb{Y}^\mu{}_\nu$,
which is obtained by a four-dimensional orthogonal transformation
that leaves unaffected the physics.
The three-dimensional part of this transformation amounts to a rotation
and hence overlaps with a Lorentz transformation,
so the two discrete Lorentz-violating solutions  
can be viewed as part of a continuous rotation-degenerate solution 
that describes the same physics as the discrete pair.

To summarize,
for case A with $\ol{\be}_4 \ne 0$ we find two possibilities
distinguished by the sign of the discriminant $D$.
Case A1 has $D \leq 0$. 
For $D<0$ it contains three Lorentz-invariant solutions 
obeying the condition \rf{caseA1}
and given by one root of the cubic \rf{cubic-la_1-eq},
while for $D=0$ only two distinct Lorentz-invariant solutions survive.
Case A2 has $D > 0$. 
It includes one Lorentz-invariant solution 
satisfying the condition \rf{caseA1}
and given by the sole real root of the cubic \rf{cubic-la_1-eq}.
This case also includes six Lorentz-violating solutions,
with the four variables $\wtnn$, $\la_1$, $\la_2$, $\la_3$ combining
in pairs and specified as roots of the quadratic \rf{quadratic-eq}.

\subsubsection{Local stability}

Next, 
we investigate the local stability of the solutions 
in the potential manifold.
To establish the local stability of unconstrained systems
it suffices to determine the eigenvalues of the hessian matrix,
which are positive definite at local minima,
negative definite at local maxima,
and indefinite at saddle points.
However,
the system of interest here is constrained,
which introduces an additional complication.
An elegant way to determine the properties of the hessian 
on the constrained surface is to work instead with the bordered hessian
\cite{rw12},
which is defined instead on an enlarged space
incorporating the Lagrange multiplier for the constraint 
along with the physical degrees of freedom.
In the present context,
the method requires first finding the determinant $\det H_B$ 
of the $4\times4$ bordered hessian 
associated with the four variables $\{\wtnn,\la_1,\la_2,\la_3\}$.
If $\det H_B<0$, 
then the hessian on the two-dimensional constrained surface
has either two positive or two negative eigenvalues.
If instead $\det H_B>0$, 
then the hessian has two eigenvalues of opposite sign,
and a principal minor must be calculated 
to determine which alternative is realized.
The principal minor is the determinant 
$\det H_{B,{\rm m}}$
of the $3\times3$ matrix obtained by removing from the hessian 
a column and a row associated with one of the variables $\la_i$.
If $\det H_{B,{\rm m}}<0$
then both eigenvalues of the constrained hessian are positive,
while if $\det H_{B,{\rm m}}>0$ then both are negative.

For case A1 with $D < 0$ and three Lorentz-invariant solutions,
we find that the determinant of the full bordered hessian $H_B$ is 
\beq
\det H_B = 
-3(\ol{\be}_2 + 2\ol{\be}_3 \la_1 + \ol{\be}_4 \la_1^2)^4 .
\label{det-bordered-hessian-LI}
\eeq
This is negative definite provided a quadratic combination 
of the variable $\la_1$ is nonzero,
\beq
\ol{\be}_2 + 2\ol{\be}_3 \la_1 + \ol{\be}_4 \la_1^2 \neq 0.
\label{quadratic-la_1}
\eeq
The zeros of this quadratic combination differ from the solutions 
when the three roots of the cubic polynomial \rf{cubic-la_1-eq}
are all distinct,
because the zeros correspond to the stationary points of the cubic 
while the extrema and saddle points of the potential correspond to its roots.
Each of the three Lorentz-invariant solutions 
therefore represents either a maximum or a minimum.
To determine which of these occurs,
we compute the principal minor
\beq
\det H_{B,\rm{m}} 
= 2(\ol{\be}_2 + 2\ol{\be}_3 \la_1 + \ol{\be}_4 \la_1^2)^3.
\label{principal-minor-LI}
\eeq
The sign of this expression matches 
the sign of the quadratic combination \rf{quadratic-la_1}.
The roots of the latter separate the three values of $\la_1$ 
corresponding to the three solutions,
so its sign alternates when they are ordered by the value of $\la_1$.
It follows that when $\ol{\be}_4 > 0$
the central solution has negative value of $\det H_{B,\rm{m}}$
and hence is a local maximum,
while the other two have positive values
with signs coinciding with that of $\ol{\be}_4$ and hence are local minima.
For $\ol{\be}_4 < 0$ the situation is reversed,
and the signs of $\det H_{B,\rm{m}}$ for the two outer extrema 
again coincide with that of $\ol{\be}_4$.

Consider next the case A2 with $D > 0$,
which has one Lorentz-invariant 
and six Lorentz-violating solutions.
The Lorentz-invariant one corresponds to the sole root of the
cubic polynomial \rf{cubic-la_1-eq}.
The determinant of the corresponding bordered hessian and the
principal minor are again given by Eqs.\ \rf{det-bordered-hessian-LI}
and \rf{principal-minor-LI}.
As this root lies outside the interval spanned by the roots of
the quadratic combination \rf{quadratic-la_1},
the sign of the principal minor is again given by the sign of
$\ol{\be}_4$.
Thus, 
if $\ol{\be}_4 > 0$ then the extremum is a local minimum,
while if $\ol{\be}_4 < 0$ it is a local maximum. 

For the Lorentz-violating solutions in case A2,
the determinant of the bordered hessian turns out to be 
\beq
\det H_B = 
\frac{D^2}{\left(\ol{\be}_3^2 - \ol{\be}_2\ol{\be}_4\right)^2}.
\label{det-bordered-hessian-LV}
\eeq
This is positive definite for $D \ne 0$.
The two eigenvalues of the constrained hessian
therefore have opposite signs, 
so the Lorentz-violating solutions 
correspond to local saddle points of the potential.

\subsubsection{Global and absolute stability}

With the extrema and their local stability properties in hand,
the issue of their global and absolute stability can be addressed.
We call an extremum globally stable
if it is locally stable
and if no locally unstable extremum
can be reached via a smooth path in field space
along which the effective potential remains finite.
This notion of global stability thus depends
on the branch structure of the potential.
The point is that two locally stable extrema 
on a single branch of the potential can in principle be connected 
via thermal fluctuations or quantum tunneling,
whereas two locally stable extrema lying on different branches
are disconnected by an infinite potential barrier. 
Also,
we refer to an extremum as absolutely stable
if it is globally stable 
and in addition lies at a lower potential
than any other globally stable extremum.
To investigate the global and absolute stability
of the various extrema,
we use a combination of analytical and graphical methods.

To set up the analytical approach,
we solve the hamiltonian constraint explicitly for one variable,
say $\la_3$,
to obtain
\beq
\la_3 = 
- \frac{\ol{\be}_1 + \ol{\be}_2 \la_1 
+ \ol{\be}_2 \la_2 + \ol{\be}_3 \la_1 \la_2}
{\ol{\be}_2 + \ol{\be}_3 \la_1 
+ \ol{\be}_3 \la_2 + \ol{\be}_4 \la_1 \la_2}.
\label{solve-la_3}
\eeq
Substituting this expression into the potential \rf{potential2}
then generates an effective potential $\ol{\mathcal{U}}(\la_1,\la_2)$
that is a function of the two remaining variables $\la_1$ and $\la_2$,
\begin{widetext}
\bea
\ol{\mathcal{U}}(\la_1,\la_2) &\equiv&
\mathcal{U}\bigl(\la_1,\la_2,\la_3(\la_1,\la_2)\bigr)
\nn\\
&&
\hskip -40pt
=\frac{\la_1\la_2\bigl(
(\ol{\be}_3^2 - \ol{\be}_2 \ol{\be}_4)\la_1\la_2 
+ (\ol{\be}_2 \ol{\be}_3 
- \ol{\be}_1 \ol{\be}_4)(\la_1 + \la_2) 
+ \ol{\be}_2^2\bigr) 
+ (\ol{\be}_2^2 - \ol{\be}_1 \ol{\be}_3) (\la_1^2 + \la_2^2) 
+ \ol{\be}_1 \ol{\be}_2(\la_1 + \la_2) 
+ \ol{\be}_1^2} {\ol{\be}_2 + \ol{\be}_3(\la_1 + \la_2) 
+ \ol{\be}_4\la_1 \la_2}.
\label{bar-U}
\eea
\end{widetext}
The crucial feature in this formula is the denominator.
Any surface in field space where it vanishes 
represents a singular surface in the definition of $\ol{\mathcal{U}}$.
Unless the numerator vanishes as well,
the behavior of $\ol{\mathcal{U}}$ across this surface
is that of a first-order pole.
This means that $\ol{\mathcal{U}}$ tends to $+\infty$
when the surface is approached from one side,
while $\ol{\mathcal{U}}$ tends to $-\infty$
when approached from the other.
The surface therefore serves as a separator 
of two distinct branches of $\ol{\mathcal{U}}$.

For case A1 with $D < 0$ and three Lorentz-invariant extrema,
we have $\la_1 = \la_2$ 
and so the denominator takes the form 
of the quadratic combination \rf{quadratic-la_1}.
The two zeros of this quadratic lie between the roots of $\la_1$ 
that define the three extrema.
We can therefore conclude that 
they lie on separate branches of the potential $\ol{\mathcal{U}}$.
This suggests that no two of the three extrema 
can be smoothly connected in field space.
However,
the above reasoning implicitly assumes
that the candidate path between the extrema 
satisfies the condition $\la_1 = \la_2$,
so the possibility remains in principle
that a more complicated path exists that avoids the singularity.

\begin{figure}
\centering
\includegraphics[width=0.4\textwidth]{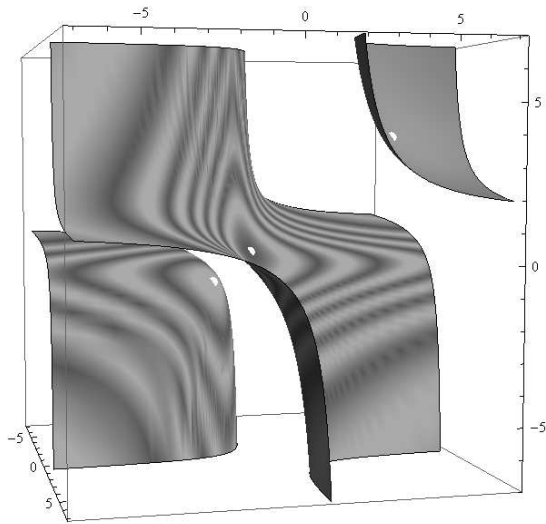}
\caption{Cubic surface for case A1,
with three Lorentz-invariant extrema.}
\label{fig1}
\end{figure}

To check this possibility,
we construct numerically a three-dimensional plot
of the cubic surface defined by the hamiltonian constraint \rf{tilde-N-eom},
using the parameters 
$\ol{\be}_1 = 0.3$, $\ol{\be}_2 = -1.3$, 
$\ol{\be}_3 = -1$, $\ol{\be}_4 = 1$.
See Fig.\ \ref{fig1}.
The $x$, $y$, $z$ axes of the plot
are labeled with values of $\la_1$, $\la_2$, $\la_3$,
respectively.
Equipotential contours of the effective potential 
$\ol{\mathcal{U}}(\la_1,\la_2)$
are displayed in grayscale shadings.
The location of the three Lorentz-invariant extrema is indicated by white dots.
Inspection of the figure reveals that the hamiltonian constraint
involves three disconnected sheets,
each containing a single stationary Lorentz-invariant extremum.
This confirms that it is impossible to transit smoothly 
from one extremum to another. 
The two extrema in case A1 that represent local minima 
are therefore both locally and globally stable.
Note,
however,
that the two globally stable extrema
generically lie at different potentials.
It follows that only the one at lower potential is absolutely stable.
Since they are separated by infinite potential barriers,
neither thermal fluctuations nor quantum tunneling between them 
can be expected to occur.

\begin{figure}
\centering
\includegraphics[width=0.4\textwidth]{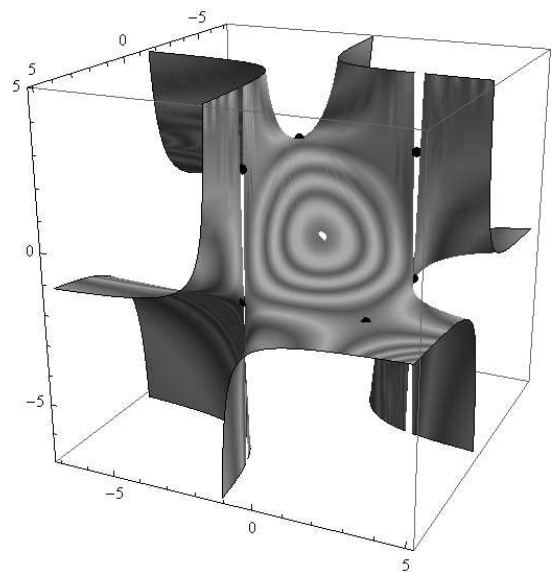}
\caption{Cubic surface for case A2 
and $\ol{\be}_3^2 - \ol{\be}_2\ol{\be}_4 < 0$,
with one Lorentz-invariant extremum and six Lorentz-violating saddle points.}
\label{fig2}
\end{figure}

For case A2 with $D > 0$ 
with one Lorentz-invariant extremum and six Lorentz-violating saddle points,
the situation depends on the sign of the combination
$\ol{\be}_3^2 - \ol{\be}_2\ol{\be}_4$.
When $\ol{\be}_3^2 - \ol{\be}_2\ol{\be}_4 < 0$,
the quadratic equation \rf{quadratic-eq} has no real roots,
so the cubic polynomial in \rf{cubic-la_1-eq} has no stationary points.
The effective potential \rf{bar-U} consists of just one branch, 
as the denominator never becomes zero.
Therefore, 
the solutions are connected by a path in the $\la_i$ space
such that the effective potential $\ol{\mathcal{U}}$ varies continuously
without passing through any singularity. 
This result is confirmed numerically in Fig.\ \ref{fig2},
which displays the cubic surface defined by the hamiltonian constraint
for the parameters
$\ol{\be}_1 = 1$, $\ol{\be}_2 = -1$,
$\ol{\be}_3 = 1$, $\ol{\be}_4 = 1$ 
using the same conventions as Fig.\ \ref{fig1}.
The Lorentz-invariant extremum is indicated by a white dot,
while the Lorentz-violating saddle points are indicated by black dots.
We thus conclude that 
the Lorentz-invariant extremum at a local mininum is globally unstable.
The excitation energy required to destabilize it
is the difference between the energies of the Lorentz-violating saddle points 
and the Lorentz-invariant local minimum.

In contrast,
when $\ol{\be}_3^2 - \ol{\be}_2\ol{\be}_4 > 0$,
the analysis is more involved.
The cubic equation \rf{cubic-la_1-eq} has two stationary points
$\la_-$, $\la_+$ corresponding to the roots 
of the quadratic combination \rf{quadratic-la_1}.
Tne single root of the cubic equation must be 
either smaller than $\la_-$ or larger the $\la_+$.
We now claim that this root for the Lorentz-invariant extremum 
and the two roots with $\la_1 = \la_2$ 
for the Lorentz-violating saddle points 
are separated by $\la_-$ and $\la_+$, 
assuming the three roots are ordered from small to large.

To check this claim,
we first evaluate the left-hand side $C$ of Eq.\ \rf{cubic-la_1-eq} 
at the centerpoint $(\la_+ + \la_-)/2$,
\beq
C = \frac{\ol{\be}_1\ol{\be}_4^2 
- 3\ol{\be}_2\ol{\be}_3\ol{\be}_4 
+ 2\ol{\be}_3^3}{\ol{\be}_4^2} .
\label{C}
\eeq
It follows that
if $\ol{\be}_4 C > 0$
then the Lorentz-invariant root $\la_{\rm LI}$ is smaller than $\la_-$,
while if $\ol{\be}_4 C < 0$ 
then $\la_{\rm LI}$ is larger than $\la_+$.
Next,
we verify that the two Lorentz-violating roots 
$\la_{{\rm LV},-}$ and $\la_{{\rm LV},+}$
given in the solutions \rf{la_LV} 
are separated from $\la_{\rm LI}$ and each other by $\la_-$ and $\la_+$.
If this is indeed the case, 
then either 
\beq
\la_{LI} < \la_- < \la_{LV,-} < \la_+ < \la_{LV,+}
\label{order1}
\eeq
or 
\beq
\la_{LV,-} < \la_- < \la_{LV,+} < \la_- < \la_{LI} .
\label{order2}
\eeq
One consistency condition for this involves the sign of the difference 
between the midpoint of $\la_{LV,-}$ and $\la_{LV,+}$.
Calculation reveals that the difference is given by 
\beq
\frac{\la_{LV,-} + \la_{LV,-}}{2} -\frac{\la_- + \la_+}{2}
= \frac{\ol{\be}_4 C}{\ol{\be}_3^2 - \ol{\be}_2\ol{\be}_4} .
\eeq
Since $\ol{\be}_3^2 - \ol{\be}_2\ol{\be}_4 > 0$ for this case,
we see that if $\ol{\be}_4 C > 0$
then the midpoint between $\la_{LV,-}$ and $\la_{LV,+}$ 
is larger than $(\la_+ + \la_-)/2$,
while if $\ol{\be}_4 C < 0$ it is smaller.
It therefore lies in the opposite direction from $\la_{LI}$
relative to $(\la_+ + \la_-)/2$,
consistent with the claim.

\begin{figure}
\centering
\includegraphics[width=0.4\textwidth]{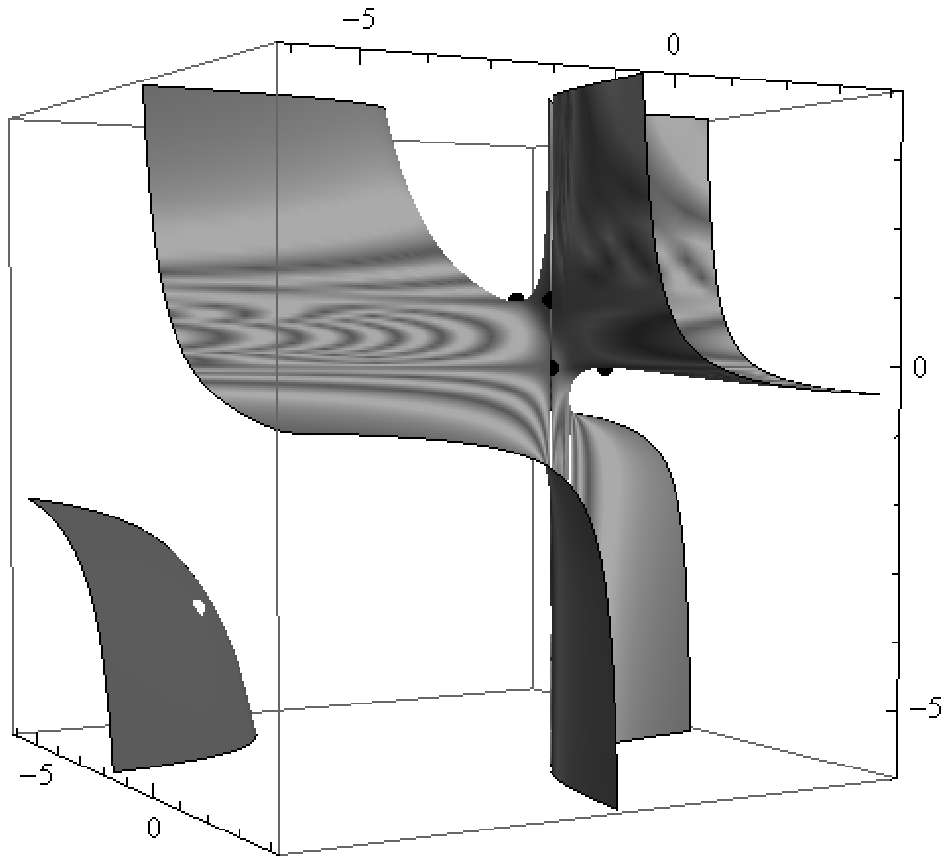}
\caption{Cubic surface for case A2 
and $\ol{\be}_3^2 - \ol{\be}_2\ol{\be}_4 > 0$,
with one Lorentz-invariant extremum and six Lorentz-violating saddle points.}
\label{fig3}
\end{figure}

To confirm that the three roots are positioned according to the claim,
we can compute explicitly the differences
\beq
\frac{\la_+ - \la_-}{2} = \frac{\sqrt{\ol{\be}_3^2 - \ol{\be}_2\ol{\be}_4}}{|\ol{\be}_4|}
\label{diff1}
\eeq
and 
\beq
\frac{\la_{LV,+} - \la_{LV,-}}{2} = 
\frac{\sqrt{D}}{2(\ol{\be}_3^2 - \ol{\be}_2\ol{\be}_4)} .
\label{diff2}
\eeq
The claim then follows if the conditions
\bea
&&
\left|\frac{\ol{\be}_1\ol{\be}_4^2 
- 3\ol{\be}_2\ol{\be}_3\ol{\be}_4 + 2\ol{\be}_3^3}
{2\ol{\be}_4(\ol{\be}_3^2 - \ol{\be}_2\ol{\be}_4)}\right|
- \frac{\sqrt{D}}{2(\ol{\be}_3^2 - \ol{\be}_2\ol{\be}_4)}
\nn\\
&&
\hskip130pt
> -\frac{\sqrt{\ol{\be}_3^2 
- \ol{\be}_2\ol{\be}_4}}{|\ol{\be}_4|} ,
\nn\\
&&
\left|\frac{\ol{\be}_1\ol{\be}_4^2 
- 3\ol{\be}_2\ol{\be}_3\ol{\be}_4 + 2\ol{\be}_3^3}
{2\ol{\be}_4(\ol{\be}_3^2 - \ol{\be}_2\ol{\be}_4)}\right|
- \frac{\sqrt{D}}{2(\ol{\be}_3^2 - \ol{\be}_2\ol{\be}_4)}
\nn\\
&&
\hskip130pt
< \frac{\sqrt{\ol{\be}_3^2 
- \ol{\be}_2\ol{\be}_4}}{|\ol{\be}_4|} ,
\nn\\
&&
\left|\frac{\ol{\be}_1\ol{\be}_4^2 
- 3\ol{\be}_2\ol{\be}_3\ol{\be}_4 + 2\ol{\be}_3^3}
{2\ol{\be}_4(\ol{\be}_3^2 - \ol{\be}_2\ol{\be}_4)}\right|
+ \frac{\sqrt{D}}{2(\ol{\be}_3^2 - \ol{\be}_2\ol{\be}_4)}
\nn\\
&&
\hskip130pt
> \frac{\sqrt{\ol{\be}_3^2 
- \ol{\be}_2\ol{\be}_4}}{|\ol{\be}_4|}
\eea
all hold.
Moving the second term in each of these relations 
to the right-hand side and squaring, 
we find that almost all terms cancel. 
All three inequalities reduce to
\beq
4(\ol{\be}_3^2 - \ol{\be}_2\ol{\be}_4)^{3/2}|\ol{\be}_4|\sqrt{D}
> 0 ,
\eeq
which is valid by inspection.
The claim that the three roots are separated 
by $\la_-$ and $\la_+$ is thus verified.

The above calculation therefore suggests
that for case A2 with $D > 0$ and $\ol{\be}_3^2 - \ol{\be}_2\ol{\be}_4 > 0$
the Lorentz-invariant stationary extremum 
and the two Lorentz-violating saddle points with $\la_1 = \la_2$ 
lie on separate branches of the effective potential
$\ol{\mathcal{U}}(\la_1,\la_2)$.
A similar argument applies for the other four Lorentz-violating saddle points.
However,
the same caveat applies here as for case A1,
as we considered only paths satisfying $\la_1 = \la_2$.
In fact,
a graphical analysis reveals that continuous paths exist 
that link the six Lorentz-violating saddle points.
This feature is manifest in Fig.\ \ref{fig3},
which plots the cubic surface defined by the hamiltonian constraint
for the parameters
$\ol{\be}_1 = 1$, $\ol{\be}_2 = 3$,
$\ol{\be}_3 = 1$, $\ol{\be}_4 = 1$,
with the conventions of Fig.\ \ref{fig1}.
The Lorentz-invariant extremum indicated by a white dot
lies on a disconnected component of the cubic surface,
while the Lorentz-violating saddle points 
indicated by black dots are clustered around the throat
of the other component.
We see that a smooth transition
from the Lorentz-invariant to the Lorentz-violating saddle points 
is impossible,
so if the Lorentz-invariant extremum is a local minimum
then it is also globally and absolutely stable.
In contrast,
smooth paths do indeed exist between any pair 
of Lorentz-violating saddle points.

\begin{figure}
\centering
\includegraphics[width=0.4\textwidth]{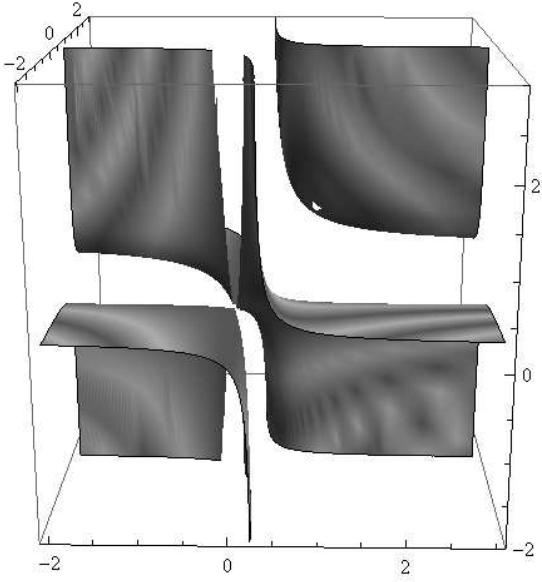}
\caption{Cubic surface for the minimal theory in Ref.\ \cite{hr11},
with one Lorentz-invariant extremum.}
\label{fig4}
\end{figure}

As a final example,
we consider the minimal theory 
obtained from the Fierz-Pauli Lagrange density
by Hassan and Rosen 
\cite{hr11}.
The corresponding the cubic surface defined by the hamiltonian constraint
is displayed in Fig.\ \ref{fig4} for the parameters
$\ol{\be}_1 = \ol{\be}_2 = 0$,
$\ol{\be}_3 = -1$, $\ol{\be}_4 = 3$,
using the conventions of Fig.\ \ref{fig1}.
These parameters yield
$D = 0$ and $\ol{\be}_3^2 - \ol{\be}_2\ol{\be}_4 =1$,
which is a limiting situation of the previous analysis.
In this case,
two sheets of the cubic surface touch at a conical singular point.
The Lorentz-invariant extremum 
$\la_0 = \la_1 = \la_2 = \la_3 = 0$ 
lies on the tip connecting the touching sheets and is unstable.
In contrast,
the Lorentz-invariant extremum 
$\la_0 = \la_1 = \la_2 = \la_3 = 1$ 
is positioned on a disconnected component of the surface,
as indicated by a white dot.
It has a negative determinant of the bordered hessian 
and a positive value for the principal minor,
so it is a local minimum that is globally and absolutely stable.

\subsection{Case B: $\ol{\be}_4 = 0$ and $\ol{\be}_3 \ne 0$}
\label{Case B}

Consider next case B with
$\ol{\be}_4 = 0$, $\ol{\be}_3 \ne 0$.
For rotationally invariant solutions with $\la_1 = \la_2 = \la_3$,
reanalysis of Eqs.\ \rf{tilde-N-eom}--\rf{la_3-la_2-eom}
reveals that the cubic \rf{cubic-la_1-eq} 
becomes replaced by the quadratic equation
\beq
\ol{\be}_1 + 3\ol{\be}_2\la_1 + 3\ol{\be}_3\la_1^2 = 0,
\label{quadratic-cubic-la_1-eq}
\eeq
which has solutions
\beq
\la_1 = \frac{-\ol{\be}_2 
\pm \sqrt{\ol{\be}_2^2-\frac{4}{3}\ol{\be}_1\ol{\be}_3}}
{2\ol{\be}_3} .
\label{la_1_be_4=0_LI}
\eeq
From the first expression in Eq.\ \rf{la_3-eom} 
it follows that $\wtnn = \la_1$,
confirming that the extrema satisfy 
the condition \rf{caseA1} for Lorentz invariance.
Note that they are real iff
$3\ol{\be}_2^2 - 4\ol{\be}_1\ol{\be}_3 \ge 0$.

To investigate local stability of these extrema,
we examine the determinant \rf{det-bordered-hessian-LI}
of the bordered hessian
and the principal minor \rf{principal-minor-LI}.
This shows that one extremum is a local maximum
of the constrained effective potential $\ol{\mathcal{U}}$
while the other is a local minimum,
depending on the sign of $\ol{\be}_2 + 2\ol{\be}_3\la_1$.

\begin{figure}
\centering
\includegraphics[width=0.4\textwidth]{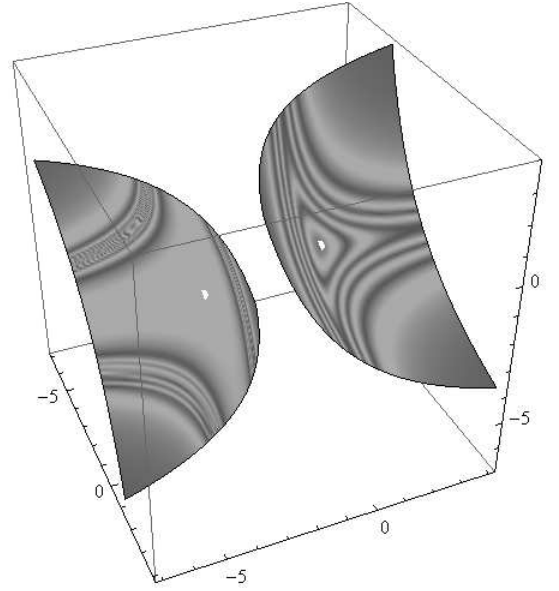}
\caption{Double-sheeted hyperboloid for case B
with two Lorentz-invariant extrema.}
\label{fig5}
\end{figure}

The effective potential in this case 
is the limit $\ol{\be}_4 \to 0$ of the expression \rf{bar-U}.
It is therefore singular along the curve 
satisfying 
\beq
\ol{\be}_2 + \ol{\be}_3(\la_1 + \la_2) = 0.
\label{singularcurve}
\eeq
The two Lorentz-invariant extrema are separated by this curve,
so we expect them to lie on separate branches of the effective potential.
This is confirmed by numerical analysis.
The double-sheeted hyperboloid defined by the hamiltonian constraint
is shown in Fig.\ \ref{fig5} for the parameters
$\ol{\be}_1 = 1$, $\ol{\be}_2 = 3$,
$\ol{\be}_3 = 1$, $\ol{\be}_4 = 0$,
using the conventions of Fig.\ \ref{fig1}.
One of the two Lorentz-invariant extrema appears on each sheet,
so they cannot be joined by a smooth curve on the surface.
The extremum that is a local minimum of the potential 
is therefore globally and absolutely stable.

For solutions without rotational symmetry,
Eqs.\ \rf{la_2-la_1-eom}--\rf{la_3-la_2-eom}
require at least two of the $\la_i$ to be equal.
Taking $\la_1 = \la_2 \ne \la_3$,
we find that $\wtnn = -\la_1 -\ol{\be}_2/\ol{\be}_3$.
Substitution into Eq.\ $\rf{la_3-eom}$ then yields
\beq
\la_1 = \la_2 
= \frac{-\ol{\be}_2 
\pm \sqrt{4\ol{\be}_1\ol{\be}_3 - 3\ol{\be}_2^2}}{2\ol{\be}_3} .
\label{la_1_be_4=0_LV}
\eeq
Note that the solutions \rf{la_1_be_4=0_LI} are real iff
$4\ol{\be}_1\ol{\be}_3 - 3\ol{\be}_2^2 \ge 0$,
contrary to the situation for the Lorentz-invariant solutions.
We also obtain 
\beq
\la_3 = \wtnn 
= \frac{-\ol{\be}_2 
\mp \sqrt{4\ol{\be}_1\ol{\be}_3 - 3\ol{\be}_2^2}}{2\ol{\be}_3} ,
\label{tilde-N_be_4=0_LV}
\eeq
confirming that the solutions obey the condition \rf{caseA2}
for Lorentz violation.
Note that the roots of the solutions for $\la_3$ and $\wtnn$ 
are interchanged relative to those for $\la_1$ and $\la_2$,
as before.
Two additional pairs of Lorentz-violating solutions are obtained
by sequential interchange of $\la_3$ with $\la_1$ and with $\la_2$.

In this case,
the determinant of the bordered hessian 
is obtained as the limit $\ol{\be}_4 \to 0$ 
of the expression \rf{det-bordered-hessian-LV},
\beq
\det H_B \to 
\bigl(3\ol{\be}_2^2 - \ol{\be}_1\ol{\be}_3\bigr)^2 .
\eeq
This is positive definite for 
$3\ol{\be}_2^2 - \ol{\be}_1\ol{\be}_3 \ne 0$,
so these solutions have a constrained hessian 
with one positive and one negative eigenvalue,
corresponding to local saddle points of the effective potential.

\begin{figure}
\centering
\includegraphics[width=0.4\textwidth]{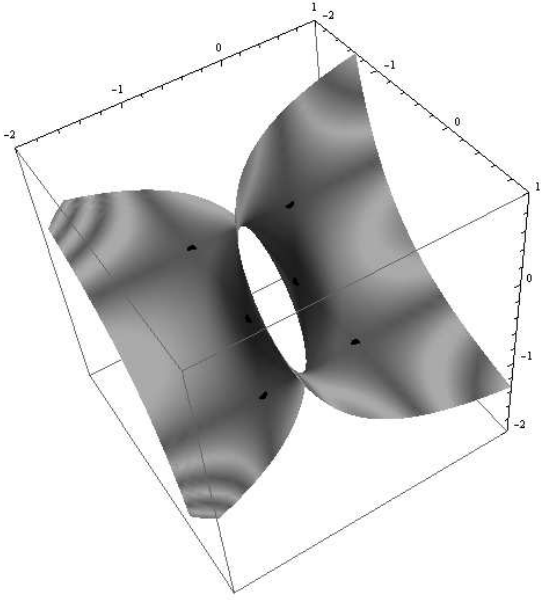}
\caption{Single-sheeted hyperboloid for case B
with six Lorentz-violating saddle points.}
\label{fig6}
\end{figure}

The Lorentz-violating saddle points 
are separated by the singular curve \rf{singularcurve}.
However,
numerical analysis reveals that all six saddle points 
lie on a unique branch of the effective potential.
The hamiltonian constraint in this case
defines a single-sheeted hyperboloid,
shown in Fig.\ \ref{fig6}
for the parameters
$\ol{\be}_1 = 1$, $\ol{\be}_2 = 1$,
$\ol{\be}_3 = 1$, $\ol{\be}_4 = 0$
with the same conventions as Fig.\ \ref{fig1}.
All six saddle points lie near the throat
of the single-sheeted hyperboloid.
Smooth transitions between all the saddle points are therefore possible.

\subsection{Case C: $\ol{\be}_4 = \ol{\be}_3 = 0$ 
and $\ol{\be}_2 \ne 0$}
\label{Case C}

\begin{figure}
\centering
\includegraphics[width=0.4\textwidth]{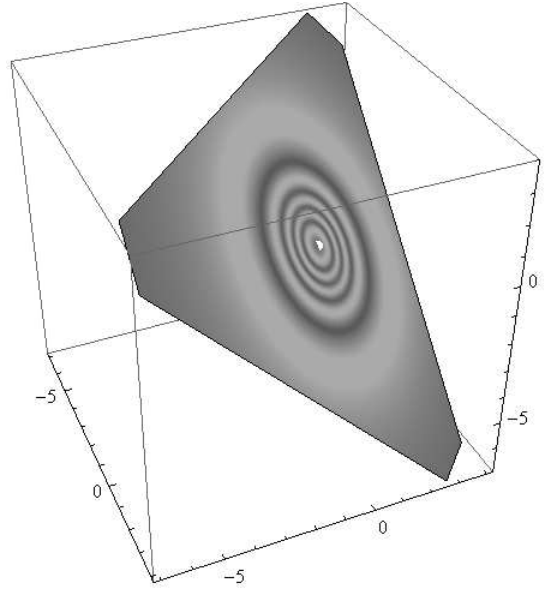}
\caption{Single-sheeted plane for case C
with a single Lorentz-invariant extremum.}
\label{fig7}
\end{figure}

Finally, 
we consider case C with $\ol{\be}_4 = \ol{\be}_3 = 0$, $\ol{\be}_2 \ne 0$.
We find that a single Lorentz-invariant solution exists, 
given by
\beq
\wtnn = \la_1 = \la_2 = \la_3 
= -\ol{\be}_1/(3\ol{\be}_2).
\eeq
This solution is a local minimum of the effective potential 
if $\ol{\be}_2 > 0$
and a local maximum if $\ol{\be}_2 < 0$.
The solution set of the hamiltonian constraint is a single-sheeted plane,
illustrated in Fig.\ \ref{fig7}
for the parameters
$\ol{\be}_1 = 1$, $\ol{\be}_2 = 3$,
$\ol{\be}_3 = \ol{\be}_4 = 0$
using the same conventions as Fig.\ \ref{fig1}.
As a result, 
if the extremum is a local minimum
then it is globally and absolutely stable.

Note that case C with a local minimum is the only scenario
for which the potential remains bounded from below
over the entire range of field variables.
Branches of the effective potential that are unbounded below
occur when the surface of the hamiltonian constraint
contains a sheet with either a local maximum or a saddle point.
This occurs for all situations in cases A and B
and for case C with a local maximum.
For these cases,
a full quantum treatment may therefore be problematic
as the path integral will probe degrees of freedom
that correspond to a potential taking arbitrarily negative values.
In this sense,
even the existence of an absolutely stable extremum
may be insufficient to guarantee stability at the quantum level
for these cases.
However,
the parameter choices $\ol{\be}_4 = \ol{\be}_3 = 0$, $\ol{\be}_2 > 0$
for case C avoid this divergence and hence
may be of particular interest in the quantum theory.
At the classical level,
in contrast,
it suffices to restrict the theory via appropriate parameter choices
to an individual sheet on which the potential is bounded from below,
so the range of viable cases is correspondingly greater.

\section{Linearized massive gravity}
\label{Linearized massive gravity}

Varying the action \rf{S} with respect to the metric $g_{\mu\nu}$
yields an equation of motion that can be written as 
\cite{hr11}
\beq
G_{\mu\nu} + \frac{m^2}{2}\sum_{n=0}^3(-1)^n\be_n
\left(g_{\mu\al}Y^\al_{(n)\nu}
+ g_{\nu\al}Y^\al_{(n)\mu}\right) = \ka T_{\mu\nu},
\label{modified-Einstein-G}
\eeq
where $T^{\mu\nu}$ is the energy-momentum tensor 
and the tensors $Y^\al_{(n)\nu}$ are given in matrix form by 
\beq
Y_{(n)}(\mathbb{X}) = \sum_{k=0}^n (-1)^k\mathbb{X}^{n-k} e_k(\mathbb{X}) .
\label{Y(n)}
\eeq
We are interested in linearizing this equation 
around Minkowski spacetime with metric $\et_{\mu\nu}$,
while allowing for deviations $\de f_{\mu\nu}$
of the fiducial metric $f_{\mu\nu}$ from $\et_{\mu\nu}$. 
We therefore define 
\beq
g_{\mu\nu} = \et_{\mu\nu} + h_{\mu\nu}, 
\quad
f_{\mu\nu} = \et_{\mu\nu} + \de f_{\mu\nu},
\label{gfexp}
\eeq
and work at first order in both $h_{\mu\nu}$ and $f_{\mu\nu}$.
To ensure the dynamical fluctuations $h_{\mu\nu}$ remain perturbative,
we take $|h_{\mu\nu}| \ll |\de f_{\mu\nu}| \ll 1$ where needed.
The deviations $\de f_{\mu\nu}$ are assumed to be constants.
Note that the presence of the background Minkowski spacetime 
implies that $\de f_{\mu\nu}$ can nonetheless produce physical effects,
as a general coordinate transformation chosen to remove $\de f_{\mu\nu}$ 
also changes the background metric to non-Minkowski form.

The square root $\mathbb{X}^\mu{}_\nu = (\sqrt{g^{-1}f}~)^\mu{}_\nu$
can be expressed in terms of $h_{\mu\nu}$ and $f_{\mu\nu}$.
The expansion
\bea
\mathbb{X}^2 &=& (\et + h)^{-1}(\et + \de f)
\nn\\
&\approx& \mathbb{1} + \et^{-1}\de f - \et^{-1}h - \et^{-1}h\,\et^{-1}\de f 
\eea
implies
\beq
\mathbb{X} \approx 
\mathbb{1} + \half\et^{-1}\de f 
- \tfrac{1}{2}\et^{-1}h + \tfrac{1}{8}\et^{-1}\de f\,\et^{-1}h 
- \tfrac{3}{8}\et^{-1}h\,\et^{-1}\de f.
\label{X}
\eeq
The $n$th product of $\mathbb{X}$ then takes the form 
\bea
\mathbb{X}^n &\approx& \mathbb{1} 
+ \frac{n}{2}\left(\et^{-1}\de f - \et^{-1}h\right)
\nn\\
&&
\hskip -10pt
- \tfrac{1}{8}\left[(n^2 - 2n)\et^{-1}\de f\,\et^{-1}h 
+ (n^2 + 2n)\et^{-1}h\,\et^{-1}\de f\right] .
\nn\\
\label{X^n}
\eea

The definition \rf{Y(n)} contains the polynomial functions $e_n(\mathbb{X})$,
which involve traces $[\mathbb{X}^n]$ of powers of $\mathbb{X}$.
Taking the trace of Eq.\ \rf{X^n} yields
\beq
[\mathbb{X}^n] \approx 4 
+ \frac{n}{2}\left([\et^{-1}\de f] - [\et^{-1}h]\right)
- \frac{n^2}{4}[\et^{-1}\de f\,\et^{-1}h] ,
\label{tr(X^n)}
\eeq
where the cyclic property of the trace has been used.
Using this expression,
we find the polynomial functions $e_n(\mathbb{X})$ take the form
\bea
e_n(\mathbb{X}) &=& \binom{4}{n} 
+ \tfrac{1}{2}\binom{3}{n-1}\left([\et^{-1}\de f] - [\et^{-1}h]\right)
\nn\\
&&
- \tfrac{1}{4}\binom{2}{n-1}[\et^{-1}\de f\,\et^{-1}h]
\nn\\
&&
- \tfrac{1}{4}\binom{2}{n-2}[\et^{-1}\de f]\,[\et^{-1}h] .
\label{e_n(X)}
\eea

Substituting the results \rf{X^n} and \rf{e_n(X)}
in the definition \rf{Y(n)}
yields the form of $Y_{(n)}(\mathbb{X})$ at first order 
in the metric and the fiducial metric.
The modified Einstein equation \rf{modified-Einstein-G} becomes
\begin{widetext}
\bea
&&
G^{\rm L}_{\mu\nu} + \frac{m^2}{2}\sum_{n=0}^3\be_n 
\biggl\{
2\binom{3}{n}\bigl(\et_{\mu\nu} 
+ h_{\mu\nu}\bigr) + \binom{2}{n-1} \left(h_{\mu\nu} - \de f_{\mu\nu} 
- \et_{\mu\nu}\bigl[\et^{-1}(h - \de f)\bigr]\right) 
\nn\\
&&
\hskip30pt
+ \left(\tfrac{1}{2}\binom{1}{n-2} + \binom{2}{n-1}\right)
\bigl[\et^{-1}\de f\bigr] h_{\mu\nu} 
+ \tfrac{1}{2}\binom{1}{n-2}\bigl[\et^{-1}h\bigr]\de f_{\mu\nu} 
\nn\\
&&
\hskip30pt
- \tfrac{1}{2}\binom{1}{n-1}\bigl[\et^{-1}\de f\,\et^{-1}h\bigr]\et_{\mu\nu}
- \tfrac{1}{2}\binom{1}{n-2}
\bigr[\et^{-1}\de f\bigr]\,\bigl[\et^{-1}h\bigl]\et_{\mu\nu}
\nn\\
&&
\hskip30pt
- \left(\tfrac{3}{4}\binom{2}{n-1} 
- \tfrac{1}{2}\binom{1}{n-1}\right)
\bigl(h\,\et^{-1}\,\de f + \de f\,\et^{-1}\,h\bigr)_{\mu\nu}
\biggr\} = \ka T_{\mu\nu} ,
\label{meeq}
\eea
\end{widetext}
where $G^{\rm L}_{\mu\nu}$ is the linearized Einstein tensor.
This result represents the desired linearization of the dynamics.

It is reasonable and usual to require that
the linearized equation \rf{meeq} is satisfied by the choice 
$h_{\mu\nu} = \de f_{\mu\nu} = T_{\mu\nu} = 0$.
This requirement constrains the parameters $\be_n$ according to 
\cite{hir12}
\beq
\sum_{n=0}^3 \binom{3}{n}\be_n 
= \be_0 + 3\be_1 + 3\be_2 + \be_3 = 0 .
\label{relation_bes}
\eeq
Introducing now a nontrivial fluctuation $h_{\mu\nu} \ne 0$ 
while maintaining $\de f_{\mu\nu} = T_{\mu\nu} = 0$,
the modified Einstein equation \rf{meeq} 
becomes
\beq
G^{\rm L}_{\mu\nu} + \frac{m^2}{2}\sum_{n=0}^3 \binom{2}{n-1}\be_n
\bigl(h_{\mu\nu} - [\et^{-1} h]\bigr) = 0 .
\label{modified-Einstein-eta3}
\eeq
If $\sum_{n=0}^3 \binom{2-1}{n}\be_n < 0$,
this equation describes tachyonic propagation
and corresponds to an unstable system.
If instead $\sum_{n=0}^3 \binom{2}{n-1}\be_n = 0$,
the mass term vanishes.
We are thus interested in the case
$\sum_{n=0}^3 \binom{2}{n-1}\be_n > 0$.
For this case,
we can rescale the parameters $\be_n \to a\be_n$
and the mass $m \to m/\sqrt{a}$
so that $\sum_{n=0}^3 \binom{2}{n-1}\be_n \to 1$,
thereby reducing the modified Einstein equation
to the Fierz-Pauli equation.
In what follows,
we can therefore take 
\beq
\sum_{n=0}^3 \binom{2}{n-1}\be_n = \be_1 + 2\be_2 + \be_3 = 1 
\label{scaling_bes}
\eeq
without loss of generality.
Note that the conditions \rf{relation_bes} and \rf{scaling_bes}
imply that only two combinations of the four parameters $\be_n$
govern the physics of the system. 
 
Suppose next that $\de f_{\mu\nu} \ne 0$.
Now Minkowski spacetime no longer solves
the modified Einstein equation \rf{meeq}
when $T_{\mu\nu} = 0$
because contributions from $\de f$ to the mass term
remain in the limit $h_{\mu\nu} \to 0$.
This means that even in the absence of matter, 
$T_{\mu\nu} = 0$,
spacetime has nonzero curvature whenever $\de f_{\mu\nu}$ is nonzero.
The presence of this curvature complicates
the form of solutions to the modified Einstein equation.
To minimize the calculational complexities
while still permitting study of relevant physical features,
we can introduce a special constant background energy-momentum tensor
chosen to cancel the $h_{\mu\nu}$-independent terms
on the left-hand side of the modified Einstein equation,
\beq
\ka T_{\mu\nu} = 
-\frac{m^2}{2}\left(\de f_{\mu\nu} - \et_{\mu\nu}[\et^{-1} \de f]\right) .
\label{tspecial}
\eeq
Note that this is conserved in Minkowski spacetime
for the spacetime-independent $\de f_{\mu\nu}$ of interest here, 
and hence all its partial derivatives vanish.
In the presence of this background,
a zero metric fluctuation $h_{\mu\nu} = 0$ solves 
the modified Einstein equation \rf{meeq},
and so spacetime is Minkowski.
Nonzero solutions for $h_{\mu\nu}$ can then be interpreted 
in analogy with standard weak-field gravitational physics in GR,
including gravitational waves in flat spacetime
and the Newton gravitational potential.

To analyze the physics of this system,
it is convenient to work in momentum space.
We can write the linearized Einstein equation in the form 
\beq
O_{\mu\nu}{}^{\al\be} h_{\al\be} = 0 
\eeq
and introduce the Fourier transform 
\beq
h_{\mu\nu}(x) = \int \frac{d^4p}{(2\pi)^4}e^{-ip\cdot x}\wth_{\mu\nu}(p) .
\eeq
This yields the momentum-space equation of motion
\beq
\wto_{\mu\nu}{}^{\al\be} \wth_{\al\be} = 0 ,
\label{Oh=0}
\eeq
where
\begin{widetext}
\bea
\wto_{\mu\nu}{}^{\al\be} &=& \left(\de_{(\mu}^\al \de_{\nu)}^\be
- \et_{\mu\nu}\et^{\al\be}\right)(p^2 + c_1 m^2)
- 2p_{(\mu} p^{(\al} \de_{\nu)}^{\be)}
+ p_\mu p_\nu\et^{\al\be} + \et_{\mu\nu} p^\al p^\be
\nn\\
&&
+ m^2\left(c_2\de_{(\mu}^{(\al}\de f_{\nu)}^{\be)} +
c_3[\et^{-1}\de f]\,\de_{(\mu}^\al\de_{\nu)}^\be +
c_4\left(\de f_{\mu\nu}
- \et_{\mu\nu}[\et^{-1}\de f]\right)\et^{\al\be}
+ c_5\et_{\mu\nu}\de f^{\al\be}\right),
\label{O}
\eea
\end{widetext}
with
\bea
c_1 &=& \sum_n \be_n \binom{2}{n-1} ,
\quad
c_2 = -\tfrac{3}{2}c_1 + \sum_n \be_n\binom{1}{n-1} ,
\nn\\
c_3 &=& \tfrac{3}{4}c_1 - \tfrac{1}{2}c_2 ,
\quad
c_4 = -\tfrac{1}{4}c_1 - \tfrac{1}{2}c_2  ,
\quad
c_5 = -\tfrac{3}{4}c_1 - \tfrac{1}{2}c_2  .
\nn\\
\eea
It is convenient to scale $m$ so that $c_1 = 1$.
The only remaining independent parameter is then $c_2$.
The first four terms in expression \rf{O} reproduce 
the usual Fierz-Pauli result
\cite{fp39},
as expected.
The tensor $\wto_{\mu\nu}{}^{\al\be}$ satisfies the symmetry properties 
\beq
\wto_{\mu\nu}{}^{\al\be} = 
\wto_{\nu\mu}{}^{\al\be} = \wto_{\mu\nu}{}^{\be\al} .
\label{symmetry-properties}
\eeq
Note, however,
that $\wto_{\mu\nu\al\be} \neq \wto_{\al\be\mu\nu}$
whenever $\de f_{\mu\nu}$ is nonzero.

\subsection{Gravitational waves}
\label{Gravitational waves}

In this section,
we consider the application of the linearized theory of massive gravity 
to the propagation of gravitational waves.
We first summarize the situation for $\de f_{\mu\nu} = 0$,
which corresponds to the Fierz-Pauli limit in Minkowski spacetime,
and then turn to the scenario with $\de f_{\mu\nu} \neq 0$. 

For $\de f_{\mu\nu} = 0$,
contracting the expression \rf{Oh=0} 
with $p^\nu$ and $\et^{\mu\nu}$ in turn
shows that it is equivalent to the conditions
\beq
\wth_{\mu\al} p^\al = 0 ,
\quad
\wth_\mu{}^\mu = 0 ,
\quad
(p^2 + m^2)\wth_{\mu\nu} = 0 .
\label{momentum-constraint}
\eeq
This set of equations represents
five constraints on the 10 components of $\wth_{\mu\nu}$
together with the usual dispersion relation $p^2 = -m^2$ 
for the five independent combinations.
The conditions and the dispersion relation are 
both particle and observer Lorentz covariant,
and the five independent combinations correspond 
to the five physical helicities of a massive spin-2 field.

The five constraint equations can be solved explicitly
by taking advantage of observer Lorentz invariance to choose
an observer frame in which the 4-momentum takes the form 
\beq
p^\mu = \bigl(E; 0, 0, p_3\bigr)
\label{fourmom}
\eeq
with $E = \sqrt{m^2 + p_3^2}$.
The particle Lorentz invariance of the system 
guarantees that the physical behavior is unchanged 
for other momentum choices in this chosen observer frame.
The five constraint equations in Eq.\ \rf{momentum-constraint} 
can then be satisfied by choosing the five independent variables as 
$\wth_{11}$, $\wth_{22}$, $\wth_{12}$, $\wth_{13}$, $\wth_{23}$
and expressing the remaining five components of $\wth_{\mu\nu}$
in terms of them,
\bea
\wth_{01} &=& \frac{p_3}{E}\wth_{13},
\quad 
\wth_{02} = \frac{p_3}{E}\wth_{23}, 
\nn\\
\wth_{03} &=& 
-\frac{E}{p_3}\wth_{00} = -\frac{p_3}{E}\wth_{33} 
= \frac{E\,p_3}{m^2}\bigl(\wth_{11} + \wth_{22}\bigr).
\label{remaining-components}
\eea

To obtain the helicity eigenstates,
we consider a rotation about the three momentum $\vec{p}$
given by the Lorentz transformation
\beq
h'_{\mu\nu} = R_\mu{}^\al R_\nu{}^\be \wth_{\al\be},
\label{rotation-h}
\eeq
with
\beq
R_\mu{}^\al = 
\begin{pmatrix}
1 & 0 & 0 & 0 \\
0 & \cos\theta & \sin\theta & 0 \\
0 & -\sin\theta & \cos\theta & 0 \\
0 & 0 & 0 & 1
\end{pmatrix} .
\label{rotation-theta}
\eeq
By definition,
a state $\ps_n$ of helicity $n$ then transforms according to 
\beq
\ps'_n = e^{in\theta} \ps_n .
\eeq
Calculation with these expressions reveals that
the helicity eigenstates of $\wth_{\mu\nu}$ 
satisfying the conditions \rf{remaining-components} are given by
\begin{align}
\wth^{(\pm 2)}_{\mu\nu} &= \frac{m^2}{2}\begin{pmatrix}
0& 0& 0& 0\\
0& 1& \pm i & 0\\
0& \pm i& -1 & 0\\
0& 0& 0& 0
\end{pmatrix},
\nn \\
\wth^{(\pm 1)}_{\mu\nu} &= \begin{pmatrix}
0& p_3& \mp i p_3& 0\\
p_3& 0& 0& -E\\
\mp i p_3& 0& 0& \pm i E\\
0& -E& \pm i E& 0
\end{pmatrix} ,
\nn \\
\wth^{(0)}_{\mu\nu} &= \begin{pmatrix}
p_3^2 & 0 & 0 & -p_3 E \\
0& -m^2/2& 0& 0\\
0& 0& -m^2/2& 0\\
-p_3E& 0& 0& E^2
\end{pmatrix}  .
\label{tilde-h-helicity-0}
\end{align}
As expected,
the five physical degrees of freedom of the massive spin-2 field
include two helicity-2 components
$\wth^{(\pm 2)}_{\mu\nu}$,
two helicity-1 components $\wth^{(\pm 1)}_{\mu\nu}$,
and a helicity-0 component $\wth^{(0)}_{\mu\nu}$.
Each component obeys the Lorentz-invariant dispersion relation
appropriate for a particle of mass $m$.

\subsubsection{Analysis in special observer frame}

Next,
we consider effects of the terms proportional to $\de f_{\mu\nu}$
in the equation of motion \rf{Oh=0}.
Selecting a special observer frame in which
the 4-momentum takes the form \rf{fourmom}
and contracting the equation of motion 
with $p^\nu$ and $\et^{\mu\nu}$ in turn,
five constraints again emerge.
Working at first order in $\de f_{\mu\nu}$,
calculation reveals they can be cast in the form
\bea
\wth_{\mu\al}p^\al &=& \tfrac{1}{3}p_\mu [\et^{-1}\de f\et^{-1}\wth]
\left(c_2 + c_4\left(1 - 2\frac{p^2}{m^2}\right)\right)
\nn\\
&&
- \tfrac{1}{2}c_2 p^\al \de f_\al{}^\be \wth_{\be\mu},
\nn\\
\wth_\al{}^\al &=& \tfrac{1}{3}[\et^{-1}\de f\et^{-1}\wth]
\left(c_2 + c_4\left(4 - 2\frac{p^2}{m^2}\right)\right) ,
\quad
\label{momentum-constraint2}
\eea
which generalizes the five constraints in Eq.\ \rf{momentum-constraint}.
Combining these results with the equation of motion \rf{Oh=0}
permits the latter to be expressed in the simplified form
\bea
&&
\bigl(p^2 + m^2(1+c_3[\et^{-1} \de f])\bigr) \wth_{\mu\nu} =
\nn\\
&&
\hskip20pt
\frac{m^2}{2}c_2\,\bigl(\wth_{\mu\al}\de f^\al{}_\nu + 
\de f_{\mu\al} \wth^\al{}_\nu\bigr)
+ c_2\,p^\al \de f_\al{}^\be \wth_{\be(\mu} p_{\nu)}
\nn\\
&&
\hskip60pt
- \frac{c_2}{3}[\et^{-1}\de f\et^{-1}\wth]
\bigl(p_\mu p_\nu + m^2\et_{\mu\nu}\bigr),
\label{dispersion-relation2}
\eea
generalizing the dispersion relation in Eq.\ \rf{momentum-constraint}.

Inspection reveals that the parameter $c_3$ 
governs an overall mass shift
set by the $\et_{\mu\nu}$-trace of the fluctuation $\de f_{\mu\nu}$ 
of the fiducial metric.
The role of the terms on the right-hand side 
of the modified dispersion relation \rf{dispersion-relation2} 
involves a nontrivial action on the components of $\wth_{\mu\nu}$,
so their physical content appears more challenging to understand.
However,
the structure of the modified dispersion relation 
implies that it can be interpreted as an eigenvalue equation,
as the combined action of the $c_2$ terms 
must be proportional to $\wth_{\mu\nu}$.
The physical content of the system
is therefore determined by the eigenvalues and eigenfunctions
of Eq.\ \rf{dispersion-relation2},
with $\wth_{\mu\nu}$ constrained 
to satisfy the conditions \rf{momentum-constraint2}.
Note that $E$ can be taken approximately equal 
to its unperturbed value $\sqrt{m^2 + p_3^2}$
in evaluating the action of the $c_2$ terms at first order in $\de f_{\mu\nu}$.

To gain insight into the physics and for simplicity,
we suppose first that $\de f_{\mu\nu}$ has the diagonal form
\beq
\de f_{\mu\nu} =
\begin{pmatrix}
a & 0 & 0 & 0\\
0 & b_1 & 0 & 0\\
0 & 0 & b_2 & 0\\
0 & 0 & 0 & b_3
\end{pmatrix}  .
\label{f-munu_diagonal}
\eeq 
Calculation then reveals five distinct eigenvalues 
for the $c_2$ terms on the right-hand side of Eq.\ \rf{dispersion-relation2},
\begin{widetext}
\bea
\label{la-12}
\la_{12} &=& -\frac{c_2}{2} m^2(b_1 + b_2),
\quad
\la_{13} = -\frac{c_2}{2} \bigl(m^2(b_1 + b_3) + p_3^2(a + b_3)\bigr),
\quad
\la_{23} = -\frac{c_2}{2} \bigl(m^2(b_2 + b_3) + p_3^2(a + b_3)\bigr)
\nn\\
\la_\pm &=& -\frac{c_2}{3}\Biggl((b_1 + b_2 + b_3)m^2 + (a + b_3)p_3^2
\pm \sqrt{\left(m^2\left(b_3 - \frac{b_1 + b_2}{2}\right) 
+ p_3^2(a + b_3)\right)^2 + \frac{3}{4}(b_1 - b_2)^2 m^4}\,\Biggr) .
\label{la-pm}
\eea
For the energy eigenvalue $p_0$,
the modified dispersion relation yields
\bea
p_0 &=& \sqrt{p_3^2 + m^2\bigl(1-c_3[\et^{-1}\de f]\bigr) + \la}
\approx
\sqrt{p_3^2 + m^2}
\left(1 - \frac{m^2 c_3[\et^{-1}\de f] 
+ \la}{2\bigl(p_3^2 + m^2\bigr)}\right),
\eea
\end{widetext}
where $\la$ takes the values \rf{la-pm}.
We thus see that the energy degeneracy 
between the five helicities in the Fierz-Pauli limit
becomes broken when $\de f_{\mu\nu} \neq 0$.

At leading order in $\de f_{\mu\nu}$,
the eigenstates corresponding to the eigenvalues 
$\la_{12}$, $\la_{13}$, $\la_{23}$ 
are the components $\wth_{12}$, $\wth_{13}$, $\wth_{23}$,
respectively.
The eigenstates corresponding to $\la_\pm$ are linear combinations of
$\wth_{11}$ and $\wth_{22}$, 
with parameters depending on $m$, $f_{\mu\nu}$, and $p_3$.
These eigenstates differ from the helicity eigenstates \rf{tilde-h-helicity-0} 
of the Fierz-Pauli spin-2 theory,
instead being nontrivial linear combinations of the latter.
The remaining components of $\wth_{\mu\nu}$
are defined by Eqs.\ \rf{remaining-components}.

The situation simplifies in the ultra-relativistic limit $p_3 \gg m$.
The energy shifts corresponding 
to the eigenvalues $\la_{13}$ and $\la_{23}$ then become equal.
The same holds for the shifts corresponding to $\la_{12}$ and $\la_-$.
As a result,
the five distinct energy eigenvalues merge into only three.
Moreover, 
the corresponding eigenstates reduce to the helicity eigenstates
\rf{tilde-h-helicity-0}.
Explicitly, 
we find
\beq
\frac{p_0}{\sqrt{p_3^2 + m^2}} \longrightarrow
\begin{cases}
1 & \text{for helicities}\pm 2,\\
1 - \tfrac 14 c_2(a + b_3) & \text{for helicities}\pm 1,\\
1 - \tfrac 13 c_2(a + b_3) & \text{for helicity }0
\end{cases}
\eeq
in the ultrarelativistic limit.

Overall,
the above treatment provides intriguing physical insight
into the behavior of the gravitational waves when $\de f_{\mu\nu} \neq 0$.
In general,
the energies of the five modes of the massive graviton undergo splitting.
This corresponds to a lifting of the degeneracy of the graviton spectrum,
and it generates `pentarefringence' or `quinquerefringence'
in propagating gravitational waves.
The pentarefringence reduces to `trirefringence' 
in the ultrarelativistic limit.

The pentarefringent behavior of massive gravitational waves
is analogous to the birefringence of electromagnetic waves
known to occur in Minkowski spacetime 
in the presence of background coefficients for Lorentz violation
\cite{km09}.
The latter effects are detectable in suitable electromagnetic experiments.
Although outside the scope of the present work,
it would be of definite interest to investigate 
the prospects of experimentally measuring 
the pentarefringence of gravitational waves
with existing and future detectors.

\renewcommand\arraystretch{1.6}
\begin{table*}
\caption{
\label{cases}
Eigenvalues and eigenfunctions for special cases.} 
\setlength{\tabcolsep}{6pt}
\begin{tabular}{lll}
\hline
\hline
Condition & Eigenvalue & Eigenfunction \\
\hline
$	C^\pm = 0	$	&	$	\half A + B	$	&	$	\dfrac{D^-}{D^+}h^{(2)} - \dfrac{A - 2B}{2D^+}h^{(1)} - 3h^{(0)}- \dfrac{A - 2B}{2D^-}h^{(-1)} + \dfrac{D^+}{D^-}h^{(2)}	$	\\	
$		$	&	$	\la_\pm \equiv \tfrac13 ({2(A + B) \pm \sqrt{(A - 2B)^2 + 12 D^+ D^-}})	$	&	$	\dfrac{D^-}{D^+}h^{(2)} + \dfrac{\la_\pm - A}{D^+}h^{(1)} - \Bigl(\dfrac{\left(B-\half A\right)(\la_\pm - A)}{D^+ D^-} + 1\Bigr)h^{(0)}	$	\\	
$		$	&	$		$	&	$	\qquad + \dfrac{\la_\pm - A}{D^-}h^{(-1)} + \dfrac{D^+}{D^-}h^{(2)}	$	\\	
$		$	&	$	\la_\pm \equiv \tfrac14({3A + 2B \pm \sqrt{(A - 2B)^2 + 12 D^+ D^-}})	$	&	$	\dfrac{D^-}{D^+}h^{(2)} - \dfrac{\la_\pm - A - B}{D^+}h^{(1)} + \dfrac{\la_\pm - A - B}{D^-}h^{(-1)} - \dfrac{D^+}{D^-}h^{(2)}	$	\\	[10pt]
$	D^\pm = 0	$	&	$	A	$	&	$	C^- h^{(2)} - C^+ h^{(-2)}	$	\\	
$		$	&	$	\frac{1}{2}A + B \pm \frac{1}{2}\sqrt{C^+ C^-}	$	&	$	C^- h^{(1)} \pm C^+ h^{(-1)}	$	\\	
$		$	&	$	\la_\pm \equiv \frac{2}{3}(A + B) \pm \frac{1}{3}\sqrt{(A - 2B)^2 + 3C^+ C^-}	$	&	$	C^- h^{(2)} + 3(A - \la_\pm)h^{(0)} + C^+ h^{(-2)}	$	\\	[10pt]
$	C^\pm, D^\pm \to 0	$	&	$	-2b	$	&	$	h^{(\pm 2)}	$	\\	
$		$	&	$	-2b - (a + b)\dfrac{|\vec{p}|^2}{m^2}	$	&	$	h^{(\pm 1)}	$	\\	
$		$	&	$	-2b - \frac{4}{3}(a + b)\dfrac{|\vec{p}|^2}{m^2}	$	&	$	h^{(0)}	$	\\	[6pt]
\hline
\hline
\end{tabular}
\end{table*}

\subsubsection{Analysis in a general helicity frame}

The treatment in the previous subsection is limited 
by the fixed value \rf{fourmom} of the momentum in the chosen frame 
and by the special form \rf{f-munu_diagonal} 
adopted for the fluctuation $\de f_{\mu\nu}$ of the fiducial metric.
Next, 
we extend the analysis and study 
the persistence of the pentarefringence effect in the general case.

Consider first an arbitrary momentum $p^\mu = \bigl(E; \vec{p}\bigr)$
in a generic observer frame,
where $E = \sqrt{m^2 + |\vec{p}|^2}$.
To describe the corresponding helicity eigenstates,
it is convenient to define the quantity 
\beq
\ol p^\mu = \bigl(|\vec{p}|;E\frac{\vec{p}}{|\vec{p}|}\,\bigr) .
\label{bar-p}
\eeq
Provided $\vec{p} \ne 0$,
which we assume to hold in what follows,
this quantity is well defined and transforms 
as the 4-momentum of a particle with mass squared $-m^2$.
We also define two other purely spacelike four-vectors,
\beq
e_1^\mu = \bigl(0; \vec{e}_1\bigr) , 
\quad
e_2^\mu = \bigl(0; \vec{e}_2\bigr),
\label{e1-e2}
\eeq
where $ \vec{e}_1$ and $\vec{e}_2$ are taken to have norm $m$,
to be mutually orthogonal,
and to be orthogonal to $\vec{p}$
such that the triple $\{\vec{e}_1,\vec{e}_2,\vec{p}\}$ 
represents a right-handed set of orthogonal basis vectors.
The four 4-vectors $p^\mu$, $\ol{p}^\mu$, $e_1^\mu$, $e_2^\mu$ 
then form a nondegenerate basis satisfying the relations
\bea
p\cdot \ol{p} &=& p\cdot e_1 = p\cdot e_2 = 0,
\nn\\
\ol{p}\cdot e_1 &=& \ol{p}\cdot e_2 = e_1\cdot e_2 = 0,
\nn\\
\ol{p}\cdot\ol{p} &=& e_1\cdot e_1 = e_2\cdot e_2 = -p^2 = m^2.
\eea
Using this basis,
the helicity eigenstates \rf{tilde-h-helicity-0}
for general momentum can be expressed as
\bea
\wth^{(\pm 2)}_{\mu\nu} &=& 
\frac{1}{2}\bigl(e_{1\,\mu}e_{1\,\nu} 
- e_{2\,\mu}e_{2\,\nu}\bigr) \mp i e_{1\,(\mu} e_{2\,\nu)},
\nn\\
\wth^{(\pm 1)}_{\mu\nu} &=& 
\bigl(e_{(\mu} \mp i e_{2\,(\mu}\bigr)\ol{p}_{\nu)},
\nn\\
\wth^{(0)}_{\mu\nu} &=& \ol{p}_\mu \ol{p}_\nu 
- \frac{1}{2}e_{1\,\mu}e_{1\,\nu} - \frac{1}{2}e_{2\,\mu}e_{2\,\nu}.
\label{tilde-h-helicity-0_gen}
\eea

With these eigenstates in hand,
we can revisit the determination of the eigenvalues and eigenvectors
of the dispersion relation \rf{dispersion-relation2},
allowing now for an arbitrary momentum $p^\mu$ 
and a general form for the fluctuation $\de f_{\mu\nu}$.
Excluding the common factor $c_2m^2/2$,
the terms on the right-hand side of the dispersion relation 
can be written as an operator $S_{\mu\nu}{}^{\al\be}$ 
acting on $\wth_{\al\be}$,
\bea
S_{\mu\nu}{}^{\al\be} \wth_{\al\be} &=&
\wth_{\mu\al}\de f^\al{}_\nu + 
\de f_{\mu\al} \wth^\al{}_\nu
+ \frac{2}{m^2} p^\al \de f_\al{}^\be \wth_{\be(\mu} p_{\nu)}
\nn\\
&&
- \frac{2}{3m^2}[\et^{-1}\de f\et^{-1}\wth]
\bigl(p_\mu p_\nu + m^2\et_{\mu\nu}\bigr).
\label{S-h}
\eea
Some algebra then reveals explicit expressions 
for the action of $S_{\mu\nu}{}^{\al\be}$ 
on the helicity eigenstates \rf{tilde-h-helicity-0_gen},
\bea
S_{\mu\nu}{}^{\al\be}\wth^{(\pm 2)}_{\al\be} &=& 
A \wth^{(\pm 2)}_{\mu\nu} 
+ D^{\mp}\wth^{(\pm 1)}_{\mu\nu}
- \frac{1}{3}\wth^{(0)}_{\mu\nu} ,
\nn\\
S_{\mu\nu}{}^{\al\be}\wth^{(\pm 1)}_{\al\be} &=& 
D^\mp\wth^{(\pm 2)}_{\mu\nu} 
+ \frac{2B + A}{2}\wth^{(\pm 1)}_{\mu\nu} 
\nn\\
&&
+ \frac{1}{3}D^\mp\wth^{(0)}_{\mu\nu} 
+ \frac{1}{2}C^\mp\wth^{(\mp 1)}_{\mu\nu} ,
\nn\\
S_{\mu\nu}{}^{\al\be}\wth^{(0)}_{\al\be} &=& 
-\frac{1}{2}C^+ \wth^{(2)}_{\mu\nu} 
+ \frac{1}{2}D^+\wth^{(1)}_{\mu\nu}
+ \frac{4B + A}{3}\wth^{(0)}_{\mu\nu} 
\nn\\
&&
+ \frac{1}{2}D^-\wth^{(-1)}_{\mu\nu}
- \frac{1}{2}C^- \wth^{(-2)}_{\mu\nu}.
\label{S-h(0)}
\eea
In these results,
the momentum-dependent dimensionless quantities $A$, $B$, $C^\pm$, $D^\pm$ 
represent linear combinations of the matrix elements of the tensor $\de f$ 
expressed in the basis spanned by
$p^\mu$, $\ol{p}^\mu$, $e_1^\mu$, $e_2^\mu$,
\bea
A &=& -\frac{1}{m^2}\bigl( e_1\cdot\de f\cdot e_1 
+ e_2\cdot\de f\cdot e_2\bigr),
\nn \\
B &=& -\frac{1}{m^2} \,\ol{p}\cdot \de f\cdot\ol{p},
\nn \\
C^\pm &=& -\frac{1}{m^2}\bigl(e_1\cdot\de f\cdot e_1 
- e_2\cdot\de f\cdot e_2 \pm 2i e_1\cdot\de f\cdot e_2\bigr),
\nn \\
D^\pm &=& -\frac{1}{m^2}\bigl(e_1\cdot\de f\cdot \ol{p} 
\pm i e_2\cdot\de f\cdot \ol{p}\bigr) .
\label{abcd}
\eea
Note that $C^- = (C^+)^*$ and $D^- = (D^+)^*$.

Using Eq.\ \rf{S-h(0)},
we find that the operator $S_{\mu\nu}{}^{\al\be}/{2m^2}$ 
has five eigenvalues that generically are distinct.
Two of them can be written as
\beq
\la_{\pm} = 
\tfrac{2}{3}(A+ B) 
\pm \tfrac 13\sqrt{(A - 2B)^2 + 3C^+ C^- + 12 D^+ D^-},
\eeq
while the other three are the roots of the cubic polynomial
\bea
&&
4\la^3 - 8(A + B)\la^2 
\nn\\
&&
\quad
+ (5A^2 + 12AB + 4B^2 - C^+ C^- - D^+ D^-)\la
\nn\\
&&
\quad
- A^3 - 4A^2B - 4AB^2 + AC^+ C^- + 2(A + 2B)D^+ D^- 
\nn\\
&&
\quad
+ C^+(D^-)^2 + C^-(D^+)^2 = 0 .
\eea
These results establish the splitting of the energies of the five modes
for the general case
and hence confirm that gravitational waves
in the theory undergo pentarefringence during propagation. 

The eigenfunctions for the five modes can also
be found by explicit calculation.
Their expressions are lengthy,
so we provide here for reference only the results for two special cases
for which the results are substantially simplified.
The first case is the scenario with $C^\pm = 0$,
while the second is the case with $D^\pm=0$. 
Table \ref{cases} displays the corresponding eigenvalues and eigenfunctions
for these two cases.

To gain further physical insight,
consider an explicit example for which the fluctuation $\de f_{\mu\nu}$ 
takes a comparatively simple form in the chosen observer frame,
\beq
\de f_{00} = a ,
\quad 
\de f_{ii} = b,
\quad i = 1,2,3 ,
\label{specialf}
\eeq
with other components of $\de f_{\mu\nu}$ being zero.
Writing the 4-momentum components as $p^\mu = (E;p_1,p_2,p_3)$
and assuming $p_2^2 + p_3^2 \ne 0$,
we can choose the basis vectors \rf{e1-e2} to be
\bea
\vec{e}_1 &=& \bigl(0,p_3,-p_2\bigr)\frac{m}{\sqrt{p_2^2 + p_3^2}},
\nn\\
\vec{e}_2 &=& \bigl(-p_2^2-p_3^2,p_1 p_2,p_1p_3\bigr)
\frac{m}{|\vec{p}|\sqrt{p_2^2 + p_3^2}} .
\eea
If $p_2^2 + p_3^2 = 0$,
then an alternative choice is possible instead,
with the physical conclusions below being unaffected.

In this example,
the linear combinations \rf{abcd} are found to have the explicit forms
\beq
A = -2 b,
\quad 
B = -b -(a + b)\frac{|\vec{p}|^2}{m^2},
\quad
C^\pm = D^\pm = 0 .
\eeq
Using the results displayed in Table \ref{cases}
for the cases $C^\pm = 0$ and $D^\pm = 0$,
we can identify the eigenvalues 
of the operator $S_{\mu\nu}{}^{\al\be}$.
The limit $C^\pm \to 0$, $D^\pm \to 0$ can be obtained with some care
from either case,
along with the corresponding eigenfunctions of $S_{\mu\nu}{}^{\al\be}$.
The results are listed in the last three rows of Table \ref{cases}.
The eigenfunctions turn out to coincide exactly 
with the helicity eigenstates.
The eigenvalues for the helicities $\pm 1$ are degenerate,
as are those for helicities $\pm 2$.

The dispersion relation \rf{dispersion-relation2} 
yields the corresponding energies as
\begin{widetext}
\beq
E = \begin{cases}
\sqrt{|\vec{p}|^2 + m^2\bigl(1 + c_3(a - 3b) - c_2 b\bigr)} 
& \text{for helicity }\pm 2,
\\
\sqrt{|\vec{p}|^2\bigl(1 - \frac{1}{2}c_2(a + b)\bigr) 
+ m^2\bigl(1 + c_3(a - 3b) - c_2 b\bigr)} 
& \text{for helicity }\pm 1,
\\
\sqrt{|\vec{p}|^2\bigl(1 - \frac{2}{3}c_2(a + b)\bigr) 
+ m^2\bigl(1 + c_3(a - 3b) - c_2 b\bigr)} 
& \text{for helicity }0,
\end{cases}
\label{E-cases}
\eeq
revealing triplet splitting.
Gravitational waves therefore experience trirefringence in this example. 
As expected,
all helicities experience a mass shift $\de m$ obeying
$\de m^2 = m^2\bigl(c_3(a - 3b) - c_2 b\bigr)$.
The helicities $\pm 1$ and 0 also undergo a shift in momentum dependence,
which modifies their group velocities.
Introducing the notation $\hat{p} = \vec{p}/|\vec{p}|$,
we find the explicit group velocities are given by 
\beq
\vec{v}_g = \frac{\partial E}{\partial \vec{p}} 
= \begin{cases}
\dfrac{\vec{p}}{E}\quad \overset{|\vec{p}|\to\infty}{\longrightarrow}\quad \hat{p} 
& \text{for helicity }\pm 2,
\\
\bigl(1-\frac{1}{2}c_2(a+b)\bigr)\dfrac{\vec{p}}{E} 
\quad\overset{|\vec{p}|\to\infty}{\longrightarrow}
\quad \sqrt{1-\frac{1}{2}c_2(a+b)}\,\hat{p} 
& \text{for helicity }\pm 1,
\\[6pt]
\bigl(1-\frac{2}{3}c_2(a+b)\bigr)\dfrac{\vec{p}}{E} 
\quad\overset{|\vec{p}|\to\infty}{\longrightarrow}
\quad \sqrt{1-\frac{2}{3}c_2(a+b)}\,\hat{p}\qquad 
& \text{for helicity }0,
\end{cases}
\label{v_g}
\eeq
\end{widetext}
where the expression for $E$ in each case 
is given by the corresponding result in Eq.\ \rf{E-cases}.

The results \rf{v_g} offer additional physical insight
into the nature of the wave propagation.
For the fluctuation $\de f_{\mu\nu}$ of the fiducial metric 
to be small relative to $\et_{\mu\nu}$
as required in the definition \rf{gfexp},
it follows that $|a|$ and $|b|$ must satisfy $|a|,|b| \ll 1$.
When the 3-momentum is also small,
$|\vec{p}| \lesssim m$,
the group velocities \rf{v_g} are then always below unity
and the 4-momenta are always timelike,
so both microcausality and energy positivity hold.
In the ultrarelativistic limit,
however,
inspection of the group velocities \rf{v_g} 
reveals that in all cases the magnitude of the group velocity obeys
\beq
|\vec{v}_g| \overset{|\vec{p}|\to\infty}{\longrightarrow} \frac{E}{|\vec{p}|} .
\label{v_g-norm}
\eeq
This implies that for $|\vec{p}| \gg m$ 
the group velocities tend from below to the values 
$1$,
$\sqrt{1 - \frac{1}{2} c_2(a + b)}$,
and $\sqrt{1 - \frac{2}{3}c_2(a + b)}$ 
for helicities $\pm 2$, $\pm 1$, and 0, 
respectively.
Therefore, 
when $c_2(a + b) > 0$
the group velocities will be subluminal for any value of the momentum, 
assuring microcausality.
However,
the unconventional momentum dependence for the helicities $\pm 1$ and 0
implies that in this case the corresponding 4-momenta 
asymptote at high $|\vec{p}|$ to a spacelike cone in 4-momentum space
and hence become spacelike,
so observer frames can be found where the energies are negative.
In constrast,
when $c_2(a + b) < 0$
the group velocities of the helicities $\pm 1$ and 0
become superluminal at sufficiently high $|\vec{p}|$
and so violate microcausality,
but the 4-momenta then asymptote to a timelike cone in 4-momentum space
and hence remain timelike in any observer frame.
This complementary behavior of microcausality and energy positivity 
is analogous to that displayed by the dispersion relation 
for a Lorentz-violating Dirac spinor with positive coefficient $c_{00}$,
as discussed in Sec.\ IV C of Ref.\ \cite{kl01}.
The superluminal features found here may parallel results 
concerning superluminal modes in ghost-free gravity 
obtained via other approaches 
\cite{ag11,diow13,frhm13}.

We can confirm the generality of the above physical interpretations
by replacing the special fluctuation \rf{specialf}
of the fiducial metric with the form 
\beq
\de f_{\mu\nu} =
\begin{pmatrix}
a & \de f_{01} & \de f_{02} & \de f_{03}\\
\de f_{01} & b_1 & 0 & 0\\
\de f_{02} & 0 & b_2 & 0\\
\de f_{03} & 0 & 0 & b_3
\end{pmatrix} .
\label{f-munu_diagonal_2}
\eeq 
This represents an extension of \rf{f-munu_diagonal}
to allow nonzero values of the components $\de f_{0i}$.
An arbitrary $\de f_{\mu\nu}$ can be converted to this form
by performing a suitable rotation of the observer frame.
Note that the fiducial metric \rf{f-munu_diagonal_2} 
with generic $\de f_{0i}$ violates 
both boost and rotational invariance even in the chosen frame,
unlike the previous example \rf{f-munu_diagonal}.

Using the expressions \rf{abcd},
the explicit forms of the parameters $A$, $B$, $C^\pm$, $D^\pm$
can be determined.
For $A$ and $B$,
we find
\bea
A &=& \frac{b_1 p_1^2 + b_2 p_2^2 + b_3 p_3^2}{|\vec{p}|^2} 
- b_1 - b_2 - b_3 ,
\nn\\
B &=& -\frac{1}{m^2}\Big(a|\vec{p}|^2 +
2\frac{f_{01} p_1 + f_{02} p_2 + f_{03} p_3}{|\vec{p}|}\sqrt{\vec{p}|^2 + m^2} 
\nn\\
&&
+ \frac{b_1 p_1^2 + b_2 p_2^2 + b_3 p_3^2}{|\vec{p}|^2}
\big(|\vec{p}|^2 + m^2\big)
\Big) .
\label{B}
\eea
The expressions for $C^\pm$ and $D^\pm$ are more involved
and are omitted here for simplicity.
The same is true for the eigenvalues and eigenfunctions,
as well as for the group velocities of the individual modes.
 
In the ultrarelativistic limit, 
however,
the situation simplifies considerably.
We find that 
the parameters $A$ and $C^\pm$ are zeroth order in large 3-momenta,
while $D^\pm$ is linear and $B$ is quadratic.
The dominant contribution to the dispersion relation \rf{dispersion-relation2}
therefore arises from the parameter $B$.
At first order in $a$ and $b$,
the dispersion relation can be written in the form 
\begin{widetext}
\beq
\frac{E}{|\vec{p}|} 
\Bigg|_{|\vec p| \to \infty}
\approx \begin{cases}
1 & \text{for helicity }\pm 2,
\\
\sqrt{1 - \dfrac{c_2}{2}\left(a + 2\dfrac{\sum f_{0i} p_i}{|\vec{p}|} 
+ \dfrac{\sum b_i p_i^2}{|\vec{p}|^2}\right)}
\qquad 
& \text{for helicity }\pm 1 ,
\\[10pt]
\sqrt{1 - \dfrac{2c_2}{3}\left(a + 2\dfrac{\sum f_{0i} p_i}{|\vec{p}|} 
+ \dfrac{\sum b_i p_i^2}{|\vec{p}|^2}\right)}
\qquad 
& \text{for helicity }0.
\end{cases}
\label{E-cases2}
\eeq
We then find the group velocities 
\beq
v_{g,i} = \frac{\partial E}{\partial p_i} 
\Bigg|_{|\vec p| \to \infty}
\approx \begin{cases}
\dfrac{p_i}{|\vec{p}|} 
& \text{for helicity }\pm 2,
\\
\dfrac{p_i\bigl(1-\frac{1}{2}c_2(a + b_i + \vec{f}_0\cdot\hat{p})\bigl) 
- \frac{1}{2}c_2 f_{0i}}{E} 
& \text{for helicity }\pm 1,
\\
\dfrac{p_i\bigl(1-\frac{2}{3}c_2(a + b_i + \vec{f}_0\cdot\hat{p})\bigl) 
- \frac{2}{3}c_2 f_{0i}}{E} 
& \text{for helicity }0 ,
\end{cases}
\label{v_g_2}
\eeq
\end{widetext}
where $\hat{p} = \vec{p}/|\vec{p}|$ as before
and $\vec{f}_0 = \bigl(f_{01},f_{02},f_{03}\bigr)$.

Note that the comparatively simple relation \rf{v_g-norm} for $|\vec{v}_g|$
still holds in all cases
at first order in $a$ and $b_i$ in the ultrarelativistic limit,
even though $\vec{v}_g$ is no longer parallel to $\vec{p}$
for the helicities $\pm 1$ and 0. 
The latter feature is typical in the linearized limit 
of theories exhibiting explicit Lorentz violation
\cite{kr10}.
It can be understood as reflecting 
the emergence of an underlying Finsler geometry
\cite{rf,bcs00},
for which the notion of distance is governed both by the metric 
and by other specified quantities.
The trajectories of massive modes 
in the presence of explicit Lorentz violation
are known to correspond to geodesics in a Finsler geometry
that can vary with helicity 
\cite{ak11}.
In the present instance,
we expect the relevant Finsler metric to be constructed 
from the metric $\et_{\mu\nu}$ 
on the approximately Minkowski spacetime
together with the fiducial metric $f_{\mu\nu}$.
Pursuing the details of this correspondence 
would be of definite interest but lies beyond our present scope. 

The results \rf{v_g-norm} and \rf{E-cases2}
suffice to examine the dependence of microcausality 
and positivity of the energy
on the generic fluctuation \rf{f-munu_diagonal_2} of the fiducial metric.
Consider,
for example,
the case with $c_2 b_1 \ge c_2 b_2 \ge c_2 b_3$
and $c_2(a + b_3) \ge 2|c_2||\vec{f}_0|$.
For this situation,
we find
\begin{widetext}
\bea
\sqrt{1 - \tfrac{1}{2}c_2(a + b_1) - |c_2||\vec{f}_0|} 
\le &|\vec{v}_g|&
\le \sqrt{1 - \tfrac{1}{2}c_2(a + b_3) + |c_2||\vec{f}_0|} 
\le 1
\qquad 
\mbox{for helicity }\pm 1 ,
\nn\\
\sqrt{1 - \tfrac{2}{3}c_2(a + b_1) - \tfrac{4}{3}|c_2||\vec{f}_0|} 
\le &|\vec{v}_g|& 
\le \sqrt{1 - \tfrac{2}{3}c_2(a + b_3) + \tfrac{4}{3}|c_2||\vec{f}_0|} 
\le 1
\qquad 
\mbox{for helicity 0}.
\eea
\end{widetext}
An argument paralleling the one given for the rotationally symmetric case
then confirms that the group velocities are subluminal for any momentum, 
so microcausality holds.
Also,
whenever the group velocities are subluminal in the ultrarelativistic limit, 
the 4-momentum becomes spacelike,
so energies turn negative in certain observer frames.
Excluding both superluminal velocities and spacelike 4-momenta
is possible ony by imposing the conditions
$b_1 = b_2 = b_3 = -a$ and $\de f_{0i} = 0$.
This implies that $\de f_{\mu\nu} \propto \et_{\mu\nu}$,
which is the only scenario in which 
the manifold Lorentz invariance remains unbroken.

\renewcommand\arraystretch{1.6}
\begin{table*}
\caption{
\label{parameters}
Parameters for the propagator.} 
\setlength{\tabcolsep}{6pt}
\begin{tabular}{lllll}
\hline
\hline
Parameter & Value & \phantom{xxxxxx} & Parameter & Value \\
\hline 
$	\aon	$	&	$	-c_2	$	&	&	$	\don	$	&	$	0	$	\\	[4pt]
$	\ato	$	&	$	\dfrac{(2c_2-3)m^2 - (2c_2+3)p^2}{12m^2}	$	&	&	$	\dtw	$	&	$	\dfrac{-(c_2-1)m^2 + p^2}{9m^4}	$	\\	[4pt]
$	\att	$	&	$	\dfrac{(2c_2-1)m^2 - (2c_2+1)p^2}{12m^2}	$	&	&	$	\dth	$	&	$	-\dfrac{c_2}{m^4}	$	\\	[4pt]
$	\atho	$	&	$	\dfrac{(4c_2+1)m^2 + (2c_2 + 1)p^2}{6m^4}	$	&	&	$	\dfo	$	&	$	2\,\dfrac{(c_2-1)m^2 - p^2}{9m^6}	$	\\	[4pt]
$	\atht	$	&	$	\dfrac{(4c_2+3)m^2 + (2c_2 + 3)p^2}{6m^4}	$	&	&	$	\dfi	$	&	$	4\,\dfrac{-(c_2-1)m^2 + p^2}{9m^6}	$	\\	[4pt]
$	\afo	$	&	$	-\dfrac{c_2(2m^2 + p^2)}{m^4}	$	&	&	$	\bon	$	&	$	\tfrac 14 (2c_2 - 3)	$	\\	[4pt]
$	\afio	$	&	$	\dfrac{(4c_2-1)m^2 - p^2}{6m^4}	$	&	&	$	\btw	$	&	$	\dfrac{-(6c_2-9)m^4 + 4(c_2-1)p^2m^2 - 4p^4}{36m^4}	$	\\	[4pt]
$	\afit	$	&	$	\dfrac{(4c_2-3)m^2 - 3p^2}{6m^4}	$	&	&	$	\bth	$	&	$	\dfrac{(2c_2-3)(2m^2 + p^2)}{2m^4}	$	\\	[4pt]
$	\asio	$	&	$	\dfrac{-(c_2-1)m^2 + p^2}{3m^6}	$	&	&	$	\bfo	$	&	$	\dfrac{ -(18c_2-21)m^4 - (8c_2-20)p^2m^2 + 8p^4}{36m^6}	$	\\	[4pt]
$	\asit	$	&	$	\dfrac{-(c_2-3)m^2 + 3p^2}{3m^6}	$	&	&	$	\bft	$	&	$	\dfrac{-(18c_2-27)m^4 - (8c_2-26)p^2m^2 + 8p^4}{36m^6}	$	\\	[4pt]
$	\aseo	$	&	$	-\dfrac{c_2}{m^2}	$	&	&	$	\bfi	$	&	$	\dfrac{(3m^2 + 2p^2)\bigl((4c_2-13)m_2 - 4p^2\bigr)}{18m^8}	$	\\	[4pt]
\hline
\hline
\end{tabular}
\end{table*}

\subsection{Propagator}
\label{Propagator}

Given an energy-momentum tensor $T^{\mu\nu}$,
the solution for the corresponding metric fluctuation $h_{\mu\nu}$
in massive gravity can be obtained in integral form
if the propagator is known.
In this section,
we determine the propagator $D_{\mu\nu}{}^{\al\be}$ 
associated with the operator $\wto_{\mu\nu}{}^{\al\be}$ in Eq.\ \rf{O},
and we use it to explore some physical features of point-mass sources.
The propagator satisfies the defining relation
\beq
D_{\mu\nu}{}^{\sigma\tau}\wto_{\sigma\tau}{}^{\al\be} 
= \de_{(\mu}^\al \de_{\nu)}^\be,
\label{relation-D-O}
\eeq
and it shares the symmetry properties \rf{symmetry-properties} of $\wto$.

Using the relations \rf{symmetry-properties} and \rf{relation-D-O},
some calculation reveals that at first order in $\de f$
the propagator can be written as 
\begin{widetext}
\bea
D_{\mu\nu}{}^{\al\be} &=&
\frac{1}{p^2+m^2}
\biggl[\de_{(\mu}^\al \de_{\nu)}^\be
- \tfrac{1}{3}\et_{\mu\nu}\et^{\al\be}
+ \frac{2}{m^2} p_{(\mu} p^{(\al} \de_{\nu)}^{\be)}
- \frac{1}{3m^2}\left(p_\mu p_\nu\et^{\al\be} + \et_{\mu\nu} p^\al p^\be
- \frac{2}{m^2}p_\mu p_\nu p^\al p^\be\right)
\nn\\
&&
\qquad{}
- \frac{m^2}{p^2+m^2}\biggl\{
\aon\,\de f_{(\mu}^\al \de_{\nu)}^\be
+ \ato\de f_{\mu\nu}\et^{\al\be} + \att \et_{\mu\nu}\de f^{\al\be}
+ \atho p_\mu p_\nu\de f^{\al\be} + \atht \de f_{\mu\nu} p^\al p^\be
+ \afo \, p_{(\mu} p^{(\al} \de f_{\nu)}^{\be)}
\nn\\
&&\qquad\qquad{}
+ \afio (\de f\cdot p)_{(\mu} p_{\nu)}\et^{\al\be}
+ \afit \et_{\mu\nu}(\de f\cdot p)^{(\al} p^{\be)}
+ \asio (\de f\cdot p)_{(\mu} p_{\nu)}p^\al p^\be
+ \asit p_\mu p_\nu(\de f\cdot p)^{(\al} p^{\be)}
\nn\\
&&
\qquad\qquad{}
+ \aseo\,(\de f\cdot p)^{(\al} p_{(\mu)}\de^{\be)}_{\nu)}
+ \aseo\,(\de f\cdot p)_{(\mu)} p^{(\al}\de^{\be)}_{\nu)}
\nn\\
&&
\qquad\qquad{}
+ \left[p\cdot\de f\cdot p\right]
\biggl(\don\,\de_{(\mu}^\al \de_{\nu)}^\be
+ \dtw\,\et_{\mu\nu}\et^{\al\be}
+ \dth\,p_{(\mu} p^{(\al} \de_{\nu)}^{\be)}
+ \dfo p_\mu p_\nu\et^{\al\be} + \dfo \et_{\mu\nu} p^\al p^\be
+ \dfi p_\mu p_\nu p^\al p^\be
\biggr)
\nn\\
&&
\qquad\qquad{}
+ \left[\et^{-1}\de f\right]
\biggl(\bon\,\de_{(\mu}^\al \de_{\nu)}^\be
+ \btw\,\et_{\mu\nu}\et^{\al\be}
+ \bth\,p_{(\mu} p^{(\al} \de_{\nu)}^{\be)}
+ \bfo p_\mu p_\nu\et^{\al\be} + \bft \et_{\mu\nu} p^\al p^\be
+ \bfi p_\mu p_\nu p^\al p^\be
\biggr)
\biggr\}\biggr] .
\nn\\
&&
\label{propagator-explicit}
\eea
\end{widetext}
This expression consists of the propagator for Fierz-Pauli massive gravity
corrected by terms linear in $\de f$.
The latter are governed by momentum-dependent parameters 
$\rh_i$ and $\si_i$,
the forms of which are displayed explicitly in Table \ref{parameters}. 
Note that all the terms with parameters $\rh_i$ are Lorentz violating,
while all the ones with $\si_i$ are Lorentz invariant.

Using the propagator \rf{propagator-explicit},
the solution to the metric for a given energy-momentum tensor $T^{\mu\nu}(x)$
is
\beq
h_{\mu\nu}(x) = 2\ka\int\frac{d^4p}{(2\pi)^4}\,e^{-ip\cdot x}
D_{\mu\nu}{}^{\al\be}\wtt_{\al\be}(p),
\label{solution_h_energy-momentum-tensor}
\eeq
where $\wtt^{\mu\nu}(p)$ is the Fourier transform of $T^{\mu\nu}(x)$,
\beq
T^{\mu\nu}(x) = 
\int \frac{d^4p}{(2\pi)^4}e^{-ip\cdot x}\wtt^{\mu\nu}(p).
\eeq
Assuming the energy-momentum tensor is conserved, 
$\prt_\mu T^{\mu\nu} = 0$,
then the solution \rf{solution_h_energy-momentum-tensor} reduces to
\begin{widetext}
\bea
h_{\mu\nu}(x) &=& 
2\ka\int\frac{d^4p}{(2\pi)^4}\,
\frac{e^{-ip\cdot x}}{p^2 + m^2 + i\ep} 
\biggl\{\wtt_{\mu\nu} 
- \tfrac{1}{3}\left(\et_{\mu\nu} + \frac{p_\mu p_\nu}{m^2}\right)\wtt
\nonumber\\
&&\qquad\quad {} 
- \frac{m^2}{p^2 + m^2 
+ i\ep}\biggl[\aon\de f_{(\mu}^\al\wtt^{\vphantom{\al}}_{\nu)\al}
+ \ato\de f_{\mu\nu}\wtt + \att\et_{\mu\nu}[\de f\cdot\tilde T]
+ \atho p_\mu p_\nu[\de f\cdot\tilde T]
+ \afio (\de f\cdot p)_{(\mu}p_{\nu)}\wtt
\nonumber\\
&&\qquad\qquad\qquad\qquad\qquad\quad{} 
+ \aseo (\de f\cdot p)^{\al}p_{(\mu}\wtt_{\nu)\al} 
+ \bigl[p\cdot\de f\cdot p\bigr]
\left(\don \wtt_{\mu\nu} + \dtw \et_{\mu\nu}\wtt 
+ \dfo p_\mu p_\nu\wtt\right) 
\nonumber\\
&&\qquad\qquad\qquad\qquad\qquad\quad{} 
+ \bigl[\et^{-1}\de f\bigr]
\left(\bon \wtt_{\mu\nu} + \btw \et_{\mu\nu}\wtt 
+ \bfo p_\mu p_\nu\wtt\right)
\biggr]\biggr\},
\label{solution-metric}
\eea
\end{widetext}
where $\wtt(p) = \wtt^\mu{}_\mu$
is the trace of the Fourier transform of the energy-momentum tensor.

As an application of the above results,
we consider the gravitational field
produced by a stationary point mass $M_1$ at the origin.
The energy-momentum tensor for this scenario is
\beq
T_1^{\mu\nu}(x) = M_1\,\de^\mu_0\de^\nu_0 \,\de^3(\vec{x}) .
\label{EM-stationary-mass}
\eeq
As required, it is conserved.
The Fourier transform is
\beq
\wtt_1^{\mu\nu}(p) = 2\pi M_1\,\de^\mu_0\de^\nu_0 \,\de(p_0) .
\label{tilde-EM-stationary-mass}
\eeq
Substitution into Eq.\ \rf{solution-metric} 
yields the solution for the metric fluctuation,
which has the structure 
\bea
h_{00}(x) &=& -\frac{8 G M_1}{3r}e^{-mr} + \mathcal{O}(\de f),
\nn\\
h_{0i}(x) &=& \mathcal{O}(\de f),
\nn\\
h_{ij}(x) &=& \frac{4 G M_1 e^{-mr}}{3 m^2 r^5}
\big[x_i x_j(m^2r^2 + 3mr + 3) 
\nn\\
&&
\hskip60pt
- \de_{ij}r^2(mr + 1)\big] + \mathcal{O}(\de f) .
\qquad
\eea
The appearance of the exponential factor $e^{-mr}$ 
is the usual Yukawa suppression arising from the graviton mass.

The $\mathcal{O}(\de f)$ contributions to the components of $h_{\mu\nu}$
can in principle be obtained by direct calculation.
However,
for our present purposes it suffices to 
deduce the gravitational potential energy $U(\vec r)$
between the point mass $M_1$ and 
a second stationary point mass $M_2$ located at $\vec r $.
The energy-momentum tensor of the second point mass
can be written as 
\beq
T_2^{\mu\nu}(x) = M_2\,\de^\mu_0\de^\nu_0 \,\de^3(\vec{x} - \vec{r}) .
\label{EM-stationary-mass2}
\eeq
Since a generic energy-momentum tensor
can be defined by variation of the matter action $S_{\rm m}$ via
\beq
T_{\mu\nu} = 
\frac{-2}{\sqrt{-g}}\frac{\de S_{\rm m}}{\de g^{\mu\nu}},
\eeq
the corresponding matter Lagrange density must take the form 
\beq
\mathcal{L}_{\rm m} \approx -\tfrac{1}{2}h^{\mu\nu} T_{\mu\nu}
\label{L-matter}
\eeq
at linear order in $h_{\mu\nu}$.
In the present context,
this represents the interaction energy between the two masses.
Since both masses are stationary,
we can directly identify $U = -\mathcal{L}_{\rm m}$.
We thereby obtain 
\bea
U(\vec{r}) &=& \int d^3x\,\half 
\bigl(h_2^{\mu\nu}(\vec{x})T^{\vphantom{\mu}}_{1,\mu\nu}(\vec{x}) 
+ h_1^{\mu\nu}(\vec{x})T^{\vphantom{\mu}}_{2,\mu\nu}(\vec{x})\bigr)
\nn\\
&=& \int d^3x\,h_1^{\mu\nu}(\vec{x})T_{2,\mu\nu}(\vec{x}) 
\nn\\
&=& M_2\, h_{1,00}(\vec{r}),
\eea
where the second equality follows by the symmetry 
under interchange of the two masses.
Obtaining an explicit expression for $U(\vec r)$
therefore requires knowledge only of the component $h_{1,00}$.
Moreover,
the $\de$ function in the energy-momentum tensor \rf{tilde-EM-stationary-mass}
implies that terms in the solution \rf{solution-metric} 
proportional to $p_\mu$ can be disregarded.
We thus find that the gravitational interaction energy $U$ 
can be written as the momentum-space integral
\begin{widetext}
\beq
U = 2\ka M_1M_2\int\frac{d^3p}{(2\pi)^3}\,
\frac{e^{-i\vec{p}\cdot\vec{r}}}{\vec{p}\,^2 + m^2}
\biggl\{\tfrac{2}{3} - \frac{m^2}{\vec{p}\,^2 + m^2}
\biggl[-(\aon + \ato + \att )\de f_{00}
+ \bigl[p\cdot\de f\cdot p\bigr](\don + \dtw)
+ \bigl[\et^{-1}\de f\bigr](\bon + \btw ) 
\biggr]\biggr\},
\label{solution-h00}
\eeq
where the various parameters and their momentum dependences
are given in Table \ref{parameters}. 
The combinations appearing in the integrand are found to be
\bea
&& 
\aon + \ato + \att =
-\frac{c_2}{3} - (c_2 + 1)\frac{\vec{p}\,^2 + m^2}{3m^2}, 
\qquad
\don + \dtw = 
-\frac{c_2}{9m^2} + \frac{\vec{p}\,^2 + m^2}{9m^4} ,
\nn\\
&& 
\bon + \btw = 
\frac{2c_2 - 3}{6} + (c_2 - 1)\frac{\vec{p}\,^2 + m^2}{9m^2} 
- \frac{(\vec{p}\,^2 + m^2)^2}{9m^4}.
\nn\\
\eea

All the momentum integrals in the expression \rf{solution-h00} 
can be evaluated explicitly
using the formulae in Appendix \ref{app:momentum-integrals}.
We find
\bea
U(\vec{r}) &=& -\frac{G M_1 M_2 e^{-mr}}{9r}\biggl[
24 - \de f_{00}\bigl((4c_2+9)mr + 8c_2 + 8\bigr)
\nn\\
&&
\hskip20pt
- \bigl(\de f_{11} + \de f_{22} + \de f_{33}\bigr)\left((4c_2-9)mr + 2c_2 + 4 
+ \frac{2}{mr} + \frac{4}{m^2r^2}\right)
\nn\\
&&
\hskip20pt
- \frac{\de f_{ij}x^ix^j}{r^2}
\left(2c_2mr + 2c_2 - 4 - \frac{6}{mr} - \frac{12}{m^2r^2}\right)
\biggr] 
+ \frac{2\ka M_1 M_2}{9m^2}\,\de^3(\vec{r})
\biggl[\de f_{00} - \tfrac{2}{3} \bigl(\de f_{11} 
+ \de f_{22} + \de f_{33}\bigr)\biggr] .
\label{h00}
\qquad
\eea
\end{widetext}
Note that $\dtw$ quadratic in momentum and $\btw$ is quartic, 
so the corresponding terms in the integral \rf{solution-h00} 
generate the ultralocal contributions to $h_{00}(\vec{r})$
proportional to $\de(\vec{r})$
that appear in the last term of this expression.

The result \rf{h00} for the gravitational potential energy 
between the two masses 
consists of the anticipated term proportional to exp$(-mr)/r$ 
for a massive particle,
together with correction terms governed by 
the diagonal components of the fluctuation $\de f_{\mu\nu}$
of the fiducial metric. 
As expected,
the term independent of $\de f_{\mu\nu}$
is scaled by 4/3 relative to the gravitational potential in GR
\cite{vvd70}.
The correction terms depend on powers $r^n$ with $n = -2,-1,0,1$,
except for the $\de$-function part.
The latter is analogous to the standard $\de$-function contribution
to the field of a dipole
\cite{jdj}.
Note that it can be ignored in considering corrections
to Newton's law between masses at distinct locations. 

Some correction terms in the gravitational potential energy \rf{h00}
are independent of the parameter $c_2$
and hence are insensitive to the form of the potential
in the action \rf{S} for massive gravity,
while others involve the product of $c_2$
with diagonal components of $\de f_{\mu\nu}$.
Moreover,
the dependences on $c_2$ and on $\de f_{\mu\nu}$ that appear in $U(\vec r)$
differ from those affecting the propagation of gravitational waves.
For example,
when $c_2 = 0$ the dispersion relation \rf{dispersion-relation2}
for gravitational waves reduces 
to the conventional Lorentz-invariant form with shifted mass parameter,
whereas the expression \rf{h00} for the gravitational potential energy
retains unconventional terms breaking Lorentz invariance. 
This suggests that observations of gravitational waves
and laboratory tests of gravity at short range
can play a complementary role in experimental searches 
for a nonzero graviton mass. 
In particular,
since the corrections to $U(\vec r)$ proportional to $\de f_{\mu\nu}$
generically break rotational invariance as well as boost invariance,
laboratory searches for anisotropic modifications 
of the Newton inverse-square law are applicable.
Recent incarnations of these experiments
have achieved impressive sensitivities to Lorentz violation
\cite{grexpts},
so establishing the implications of short-range tests 
in the current context would be of definite interest. 

The manifold Lorentz invariance is preserved in the special scenario
with the fiducial metric proportional to the Minkowski metric, 
\beq
\de f_{\mu\nu} = \ep \et_{\mu\nu} ,
\label{de-f=ep.eta}
\eeq
with $\ep$ a perturbative constant.
The gravitational potential energy \rf{h00} then simplifies to
\bea
U(\vec{r})\Bigr|_{\de f = \ep\et} &=& 
-\frac{4G M_1 M_2e^{-mr}}{3r}\bigl(2 + \ep(3 - c_2)mr\bigr)
\nn\\
&&+ \frac{2\ka M_1 M_2}{3m^2}\de^3(\vec{r})\, \ep .
\label{h00-special}
\eea
At first order in $\ep$,
the factor correcting the usual exponential term 
can be absorbed in a mass shift $m \to m + \de m$,
\bea
U(\vec{r})\Bigr|_{\de f = \ep\et} &=& 
-\frac{8G M_1 M_2e^{-(m + \de m)r}}{3r}
\nn\\
&&
+ \frac{2\ka M_1 M_2}{3m^2}\de^3(\vec{r})\, \ep ,
\label{h00-special2}
\eea
where
\beq
\de m = \ep\,\frac{c_2 - 3}{2}\,m .
\label{de-m}
\eeq
It is interesting to compare this result
with the parallel analysis for gravitational waves.
Substituting the special fluctuation \rf{de-f=ep.eta}
into the constraints \rf{momentum-constraint2} 
and the general dispersion relation \rf{dispersion-relation2}
yields
\bea
\wth_{\mu\al}{}^\al &=& \wth_\al{}^\al = 0 ,
\nn\\
(p^2 + m^2)\wth_{\mu\nu} &=& 
\ep m^2(c_2 + 4c_3)\wth_{\mu\nu}
\nn\\
&=& \ep m^2(3 - c_2)\wth_{\mu\nu} .
\label{dispersion-relation-shifted-mass}
\eea
We see that the unperturbed constraints \rf{momentum-constraint}
are recovered.
Also,
the dispersion relation can be written as
\beq
\bigl(p^2 + (m + \de m)^2\bigr)\wth_{\mu\nu} = 0,
\eeq
with $\de m$ given by Eq.\ \rf{de-m}.
The special choice \rf{de-f=ep.eta} for the fiducial metric
is thus seen to produce the same mass shift 
both in Newton gravity and in gravitational waves.

\section{Summary}
\label{Summary}

In this work,
we investigate the role of Lorentz symmetry 
in ghost-free massive gravity.
Both Lorentz-invariant and Lorentz-violating solutions of the potential 
are determined and their local and global stability are established.
The propagation of gravitational waves and the Newton limit 
of the theory are studied in approximately Minkowski spacetime. 

The main body of the paper begins in Sec.\ \ref{Setup}
with the staging for the subsequent derivations.
The action $S$ for massive gravity is provided in Eq.\ \rf{S},
and the potential is expressed using a matrix $\mathbb{Y}^\mu{}_\nu$ 
that is well suited for calculation.
The spacetime symmetries of the action $S$ and some of their features
are described in Sec.\ \ref{Spacetime symmetries},
using key transformations including 
general coordinate transformations,
local Lorentz transformations,
diffeomorphisms,
manifold Lorentz transformations,
and the CPT transformation.
The decomposition of the matrix $\mathbb{Y}^\mu{}_\nu$ 
using convenient lapse and shift variables
for calculational purposes is presented 
in Sec.\ \ref{Matrix decomposition}.

The extrema and saddle points of the potential 
for ghost-free massive gravity are the focus of 
Sec.\ \ref{Static extrema}.
In terms of the four field variables 
$\wtnn$, $\la_1$, $\la_2$, $\la_3$
and the four parameters 
$\ol{\be}_1$, $\ol{\be}_2$, $\ol{\be}_3$, $\ol{\be}_4$,
the solutions are governed by the potential 
$\mathcal{U}(\mathbb{Y})$ given in Eq.\ \rf{potential2}.
We explicitly determine and classify the solutions 
with vanishing curvatures for the dynamical and fiducial metrics,
proving that they are either Lorentz invariant
or break four of the six generators of the Lorentz group. 
The technique of the bordered hessian is adopted
to investigate local stability of the solutions,
revealing that the Lorentz-invariant ones
are either local minima or maxima
while the Lorentz-violating ones are saddle points.
To explore the issue of global stability,
the branch structure of the potential is studied.
Using a combination of analytical and numerical methods,
we determine the sheet structures of the surfaces
defined by the hamiltonian constraint \rf{tilde-N-eom}
and the corresponding forms of the potential.
This verifies that special values of the parameters $\ol{\be}_i$
allow the existence of locally, globally, and absolutely stable extrema.

The linearized limit of the equations of motion for massive gravity
is studied in Sec.\ \ref{Linearized massive gravity}.
Allowing for small deviations of the dynamical metric $g_{\mu\nu}$
and the flat nondynamical fiducial metric $f_{\mu\nu}$
from the Minkowski metric $\et_{\mu\nu}$,
we obtain the modified Einstein equation \rf{meeq}.
The momentum-space equation of motion is constructed for the scenario 
describing excitations of the fluctuation $h_{\mu\nu}$
of the dynamical metric in a Minkowski spacetime. 
One application is to the propagation of gravitational waves.
Working first in a special observer frame
with a diagonal form for the fluctuation $\de f_{\mu\nu}$
of the fiducial metric,
we find the energies of the five propagating modes.
This reveals that for nonzero $\de f_{\mu\nu}$
the gravitational waves experience pentarefringence,
reducing to trirefringence in the ultrarelativistic limit.
Generalizing the analysis to an arbitrary helicity frame
verifies these results
and shows the group velocities of the gravitational-wave modes 
can include superluminal and subluminal components.
For the subluminal case,
the mode energies become negative in certain observer frames.
These results match typical behaviors in other Lorentz-violating theories.

Section \ref{Linearized massive gravity} also contains
an investigation of the Newton limit.
We determine the propagator \rf{propagator-explicit}
for massive gravity in the linearized limit,
demonstrating that it can be written as
the Fierz-Pauli propagator corrected by terms linear in $\de f_{\mu\nu}$.
As an explicit example,
we construct the solution for the fluctuation $h_{\mu\nu}$
generated by a stationary point mass
and determine the gravitational potential energy $U$
between two point masses separated by a distance $r$.
Some integrals useful for this derivation are presented
in Appendix \ref{app:momentum-integrals}.
The result \rf{h00} for $U$
is the usual Fierz-Pauli potential of the Yukawa form 
modified by terms linear in $\de f_{\mu\nu}$.
The dependence on $\de f_{\mu\nu}$ differs from the one 
affecting gravitational waves,
suggesting that experiments in the two regimes
can provide complementary measures of the physics of massive gravity.

The results in this paper provide a guide 
to choices of parameters in the potential for massive gravity 
that guarantee local, global, and absolute stability of extrema of the action.
They also reveal that non-Minkowski fiducial metrics
generate physical effects from Lorentz violation
that could be observable in measurements of gravitational waves
and in searches for short-range modifications of Newton gravity.
The theoretical and phenomenological results obtained here 
provide directions for future works seeking insights
into the physics of this remarkable subject.

\section*{Acknowledgments}

This work was supported in part
by the U.S.\ Department of Energy under grant {DE}-SC0010120,
by the Portuguese Funda\c c\~ao para a Ci\^encia e a Tecnologia
under grants SFRH/BSAB/150324/2019 and UID/FIS/00099/2019,
and by the Indiana University Center for Spacetime Symmetries.

\appendix

\begin{widetext}
\section{Momentum integrals}
\label{app:momentum-integrals}

This appendix presents various euclidean-space momentum integrals
used in the calculation of the gravitational potential energy \rf{h00} 
in Sec.\ \ref{Propagator}.
The two elementary integrals used in the derivation
are
\beq
I_m(r) \equiv 
\int d^3p\,\frac{e^{-i \vec{p}\cdot\vec{r}}}{\vec{p}\,^2 + m^2} = 
\frac{2\pi^2}{r}e^{-mr}
\label{integral1}
\eeq
and
\beq
\int d^3p\,e^{-i \vec{p}\cdot\vec{r}} = (2\pi)^3\de^3(\vec{r}) .
\label{integral2}
\eeq
The other integrals required can be expressed in terms of these.
We find 
\bea
&&
\int d^3p\,\frac{e^{-i \vec{p}\cdot\vec{r}}}{\bigl(\vec{p}\,^2 + m^2\bigr)^2} 
= I_m(r) \frac{r}{2m} ,
\qquad
\int d^3p\,\frac{e^{-i \vec{p}\cdot\vec{r}}}
{\bigl(\vec{p}\,^2 + m^2\bigr)^2}\,p_i p_j = 
\tfrac{1}{2}I_m(r)\left( \delta_{ij} - \frac{1 + mr}{r^2}\,x_i x_j\right),
\nn\\
&&
\int d^3p\,\frac{e^{-i \vec{p}\cdot\vec{r}}}{\vec{p}\,^2 + m^2}\,p_i p_j =
\frac{(2\pi)^3}{3}\delta^3(\vec{r})\delta_{ij}
+ I_m(r)\left(\frac{2 + mr}{2r^2} 
- \frac{6 + 3mr + 2m^2 r^2}{2r^4} x_i x_j \right).
\nn\\
\label{integral5}
\eea

Note that some of these integrals lack absolute convergence
and so require care in evaluation.
For example,
naive calculation of the first integral in Eq.\ \rf{integral5}
by applying spherical coordinates,
integrating over angles,
and then integrating over the modulus of the momentum 
produces an erroneous result.
The correct expression can be derived 
by adapting the techniques in Refs.\ \cite{blp82,ga13},
\end{widetext}
\bea
&&
\hskip-40pt
\int d^3p\, e^{-i \vec{p}\cdot\vec{r}}
\frac{p_i p_j}{\bigl(\vec{p}\,^2 + m^2\bigr)^2}
\nn\\
&&
= \tfrac{1}{2}i \partial_i \int d^3p\, 
e^{-i \vec{p}\cdot\vec{r}} 
\left(-\tfrac 12 \frac{\partial}{\partial p_j}\right)
\frac{1}{\vec{p}\,^2 + m^2} 
\nn\\
&&
= i\partial_i \int d^3p \left(\tfrac 12\frac{\partial}{\partial p_j} 
e^{-i \vec{p}\cdot\vec{r}}\right)
\frac{1}{\vec{p}\,^2 + m^2} 
\nn\\
&&
= \tfrac{1}{2}\partial_i\left(x_j\int d^3p \,
\frac{e^{-i \vec{p}\cdot\vec{r}}}{\vec{p}\,^2 + m^2}\right) 
\nn\\
&&
= \tfrac{1}{2}\partial_i\bigl(x_j I_m(r)\bigr) 
\nn\\
&&
= \tfrac{1}{2}I_m(r)\left( \de_{ij} - (1 + mr)\frac{x_i x_j}{r^2}\right),
\eea
whereupon contracting with $x_i x_j/r^2$ yields the quoted result.

To obtain the last integral in Eq.\ \rf{integral5},
we apply $\partial_i \partial_j$ to both sides of Eq.\ \rf{integral1}.
Contracting the result with $x_i x_j/r^2$ 
yields the claimed result for $\vec{r} \not= 0$.
The result at the origin of $\vec{r}$ is nontrivial
because the integrand remains finite at large $|\vec{p}|$
and hence generates an extra term involving $\de^3(\vec{r})$. 
To fix this term,
it suffices to adopt the ansatz
\cite{ga13}
\beq
\int d^3p\, e^{-i \vec{p}\cdot\vec{r}}\frac{p_i p_j}{\vec{p}\,^2 + m^2} =
a(r)\de_{ij} + b(r)x_i x_j + c\de^3(\vec{r})\de_{ij},
\eeq
where $a(r)$ and $b(r)$ are functions of $r$ and $c$ is constant. 
Both sides of this equation are symmetric two-tensors 
under the rotation group.
Contracting with $\de^{ij}$ yields an integral 
that can be determined via the expressions \rf{integral1} and \rf{integral2},
which fixes the constant $c$ and the combination $3a(r) + b(r)r^2$.
Contracting with $x_i x_j/r^2$ 
yields the combination $a(r) + b(r)r^2$, 
establishing the desired result.

\end{document}